\documentclass[sigplan,10pt]{acmart}
\usepackage{prp-macros}
\renewcommand\footnotetextcopyrightpermission[1]{}
\usepackage{tikz}
\usepackage{amsmath}
\usepackage{mathtools}
\usepackage[linesnumbered,ruled,vlined]{algorithm2e}
\usepackage{ulem}
\usepackage{xfrac}

\newcommand{\sys}{Quiver\xspace}

\usepackage[font={small, bf, sf}, textfont={small, sf}]{caption}
\usepackage{float}

\usepackage[small, compact]{titlesec}

\usepackage{filecontents}

\graphicspath{{..}}

\newcommand*{\circled}[1]{\lower.7ex\hbox{\tikz\draw (0pt, 0pt)%
    circle (.5em) node {\makebox[1em][c]{\small #1}};}}

\begin{document}

\title{\sys: Supporting GPUs for Low-Latency, High-Throughput GNN Serving with \protect\\ Workload Awareness}

\author[]{Zeyuan Tan*$^{1}$ \quad Xiulong Yuan*$^{1}$ \quad Congjie He*$^{1}$}
\author[]{Man-Kit Sit$^{1}$ \quad Guo Li$^{2}$ \quad Xiaoze Liu$^{1}$ \quad Baole Ai$^{3}$ \quad Kai Zeng$^{3}$ \quad Peter Pietzuch$^{2}$ \quad Luo Mai$^{1}$}

\author[]{\hspace{36pt} }

\author[]{$^{1}$University of Edinburgh \quad $^{2}$Imperial College London \quad $^{3}$Alibaba Group}

\begin{abstract}
Systems for serving inference requests on graph neural networks~(GNN) must combine low latency with high throughout, but they face irregular computation due to skew in the number of sampled graph nodes and aggregated GNN features. This makes it challenging to exploit GPUs effectively: using GPUs to sample only a few graph nodes yields lower performance than CPU-based sampling; and aggregating many features exhibits high data movement costs between GPUs and CPUs. Therefore, current GNN serving systems use CPUs for graph sampling and feature aggregation, limiting throughput. 

We describe \emph{\sys}, a distributed GPU-based GNN serving system with low-latency and high-throughput. \sys's key idea is to exploit \emph{workload metrics} for predicting the irregular computation of GNN requests, and governing the use of GPUs for graph sampling and feature aggregation: (1)~for graph sampling, \sys calculates the \emph{probabilistic sampled graph size}, a metric that predicts the degree of parallelism in graph sampling. \sys uses this metric to assign sampling tasks to GPUs only when the performance gains surpass CPU-based sampling; and (2)~for feature aggregation, \sys relies on the \emph{feature access probability} to decide which features to partition and replicate across a distributed GPU NUMA topology. We show that \sys achieves up to 35$\times$ lower latency with a 8$\times$ higher throughput compared to state-of-the-art GNN approaches (DGL and PyG). 

\let\thefootnote\relax\footnotetext{* Co-first authors.} \let\thefootnote\relax\footnotetext{Xiulong Yuan (Tsinghua University) and Xiaoze Liu (Zhejiang University) worked on Quiver while they were visiting the University of Edinburgh.}
\end{abstract}

\settopmatter{printfolios=true}
\settopmatter{printacmref=false}

\maketitle

\pagestyle{plain}
 

\section{Introduction}
\label{sec:intro}

Many internet, financial, and scientific applications rely on serving inference requests on graph neural networks~(GNNs): examples include real-time fraud detection~\cite{10.1145/3511808.3557136,xie2021graph}, cyber-attack prevention~\cite{NEURIPS2020_690d8398}, product recommendations~\cite{recommendation1, pinsage}, complex dataset analysis~\cite{graphdl}, and particle simulations~\cite{sanchez2020learning}. 

When receiving a GNN inference request from an application, a GNN serving system samples the neighborhood within a graph. It begins at a seed node, aggregates the features associated with multiple levels of neighboring nodes, and passes the aggregated feature tensors to a deep neural network~(DNN) for the inference computation. Feature tensors are often large, because they may constitute multi-modal data such as images and text~\cite{krishnamurthy2023deep, addanki2021large}. If feature tensors exceed the capacity of a single server, they must be partitioned across servers.

To support large-scale applications with many concurrent inference requests, GNN serving systems must combine low latency with high throughput. This is challenging due to the \emph{irregular computation} that GNN serving exhibits: it typically involves large graphs with hundreds of millions of nodes and edges~\cite{addanki2021large, hamilton2017inductive} that have a high degree of skew~\cite{kondor2008skew}, \ie a proportion of graph nodes have significantly more neighbors than others. When performing multi-level neighbor sampling on these graphs for different serving requests, there is a considerable variation in the number of sampled graph nodes (from hundreds to millions), leading to variance in the aggregated feature size (from MBs to GBs).

For example, sampling the Reddit graph~\cite{hamilton2017inductive}, a typical internet graph, for a batch of 1,000~requests can yield anything from 4,000 to 300,000~neighbors, with feature tensors ranging from 5\unit{MB} to 7\unit{GB}. When a system ingests hundreds of thousands of inference requests per second~(typical for recommender systems~\cite{fan2019graph} and fraud detection~\cite{xu2021towards}), they must sample 10s of millions of graph nodes and aggregate 100s of GBs of feature data.

Due to this irregular computation pattern, current GNN systems (DGL~\cite{dgl}, PyG~\cite{pyg}, AliGraph~\cite{aligraph} and others~\cite{gnnlab, bgl, p3}) use CPUs for graph sampling and feature aggregation, only relying on GPU acceleration for DNN inference. While this reduces latency under different computational loads, it limits throughput: \eg with a latency target of below 30\unit{ms}, DGL can only handle a few 1000s of requests per second.

While GPU-based sampling implementations have been proposed~\cite{nextdoor, torch-direct}, they lead to unpredictable latencies: GPU-based graph sampling is slower than its CPU counterpart when processing requests that return fewer than 1,000 neighbours or use a small request batch size below hundreds~\cite{clipper,gao2018low}. This means that any system design that statically decides to use GPUs for sampling suffers from latency spikes. In addition, feature aggregation leads to a large amount of data movement, which causes GPUs to be bottlenecked: aggregating features on a larget real-world graph moves 100s of GBs of data per second, thus exhausting PCIe bandwidth.



Our goal is to explore a new design for a GNN serving system that exploits GPUs for graph sampling and feature aggregation for high throughout while meeting stringent latency goals.  Our key idea is for the system to take workload properties of the GNN requests into account when allocating computation to resources. More specifically, the system obtains easily computable \emph{workload metrics} about the associated graph data at runtime, which lets it decide (i)~when to allocate sampling tasks in a GNN request batch to GPUs and (ii)~how to place features across GPUs to avoid communication bottlenecks.








We describe \textbf{\sys}, a distributed GPU-based GNN serving system that leverages workload metrics when processing requests with low latency while achieving high throughput. To serve GNN inference requests, \sys is given a graph with features, a sampling method, and a DNN. It replicates this graph and partition its features on distributed servers. \sys then executes graph sampling, feature aggregation, and DNN inference as computational tasks on GPUs and CPUs in a streaming pipeline. It does this in a workload aware fashion by making the following contributions:

\mypar{(1)~Workload-aware GNN sampling} To account for the irregular computation of GNN sampling tasks, \sys dynamically schedules sampling tasks onto GPUs and CPUs based on a novel workload metric: \emph{probabilistic sampled graph size}~(PSGS). PSGS is an estimate of the sampled neighborhood size, and thus the computational load of a given sampling task. With a large PSGS, the sampling computation benefits from being scheduled on GPUs; with a small PSGS, sampling is completed more quickly on CPUs.

To obtain PSGS, \sys calculates the probability of sampling the neighbors of each seed node in the graph, extends the probabilities to multi-hop neighbors, and aggregates them by combining all possible sampling paths.

%


When executing GNN requests, \sys batches requests and considers the PSGS estimates of different batch sizes and the associated confidence intervals. To make the scheduling decisions robust, it assigns the batch size with the highest confidence to GPUs or CPUs based on the PSGS estimate. 




\mypar{(2)~Workload-aware GNN feature placement} \sys decides on the assignment of feature tensors to GPUs based on another novel workload metric: \emph{feature access probability}~(FAP). FAP predicts the likelihood of a feature being accessed when sampled as part of a multi-hop neighbor. \sys uses FAP to determine which features to place close to particular GPUs while fully utilizing NVLink and InfiniBand.

To obtain FAP, \sys calculates, for each feature, the probability that a node is sampled as a one-hop neighbor, extends the probabilities to be sampled as a multi-hop neighbor, and aggregates them when multiple neighbors are chosen as seed nodes in a request batch.



The presence of NVLink and InfiniBand on GPU servers significantly reduce the latency when fetching features. Therefore, \sys considers the \emph{GPU NUMA topology}, in addition to FAP, for feature placement, balancing partitioning and replication on GPU servers: without NVLink, \sys replicates popular features on all GPUs, avoiding data fetches over PCIe; with NVLink, which can provide 600\unit{GB/s} between GPUs, \sys places more features on GPUs by partitioning (instead of replicating) popular features.


To reduce the latency of feature aggregation, \sys uses \emph{one-sided reads} to retrieve features: it bypasses CPUs, which can become a bottleneck when coordinating a large number of features to move to GPUs, and launches data movement calls directly from GPU kernels. This also allows \sys to fully utilize the high bandwidth of NVLink and InfiniBand.

\tinyskip

\noindent
Our experiments show that \sys outperforms state-of-the-art GNN systems (PyG~\cite{pyg}, AliGraph~\cite{aligraph}, DGL~\cite{dgl}) when serving 6~GNN models from the OGB benchmarks~\cite{hu2020open}. When the serving cluster is overloaded, \sys still achieves this latency threshold, while the baseline latency rises to over 1,000\unit{ms}. \sys maintain low latency performance while the number of servers is increased. For the MAG240M graph dataset, \sys achieves up to 6$\times$ higher throughput than DistDGL~\cite{distdgl} and P3~\cite{p3}.

\sys is available as open-source~\footnote{\url{https://github.com/quiver-team/torch-quiver}}, and its techniques for workload awareness have seen adoption in industry GNN serving systems~\cite{pyg, dgl, graphlearn}.

\section{Latency and Throughput in GNN Serving}
\label{sec:motivation}

We provide background on GNN serving and the challenges in achieving low-latency, high-throughput processing. We discuss the limitations of existing system designs and introduce our goals for a low-latency GNN serving system.

\subsection{GNN serving}
\label{sec:motivation:gnn}

GNN serving is used as part of many applications, \eg recommender systems~\cite{ekko2022} in which GNNs resolve the cold start issue for recommendations; fraud detection systems~\cite{xu2021towards} in which GNNs detect long-range dependencies between transactions; smart transport~\cite{traffic} in which GNNs optimize recommended routes; and applications in science~\cite{pointcloud}, \eg by using GNNs to predict the positions of particles over time in particle simulations.

\begin{figure}[t]
  \centering
  \includegraphics[width=\linewidth]{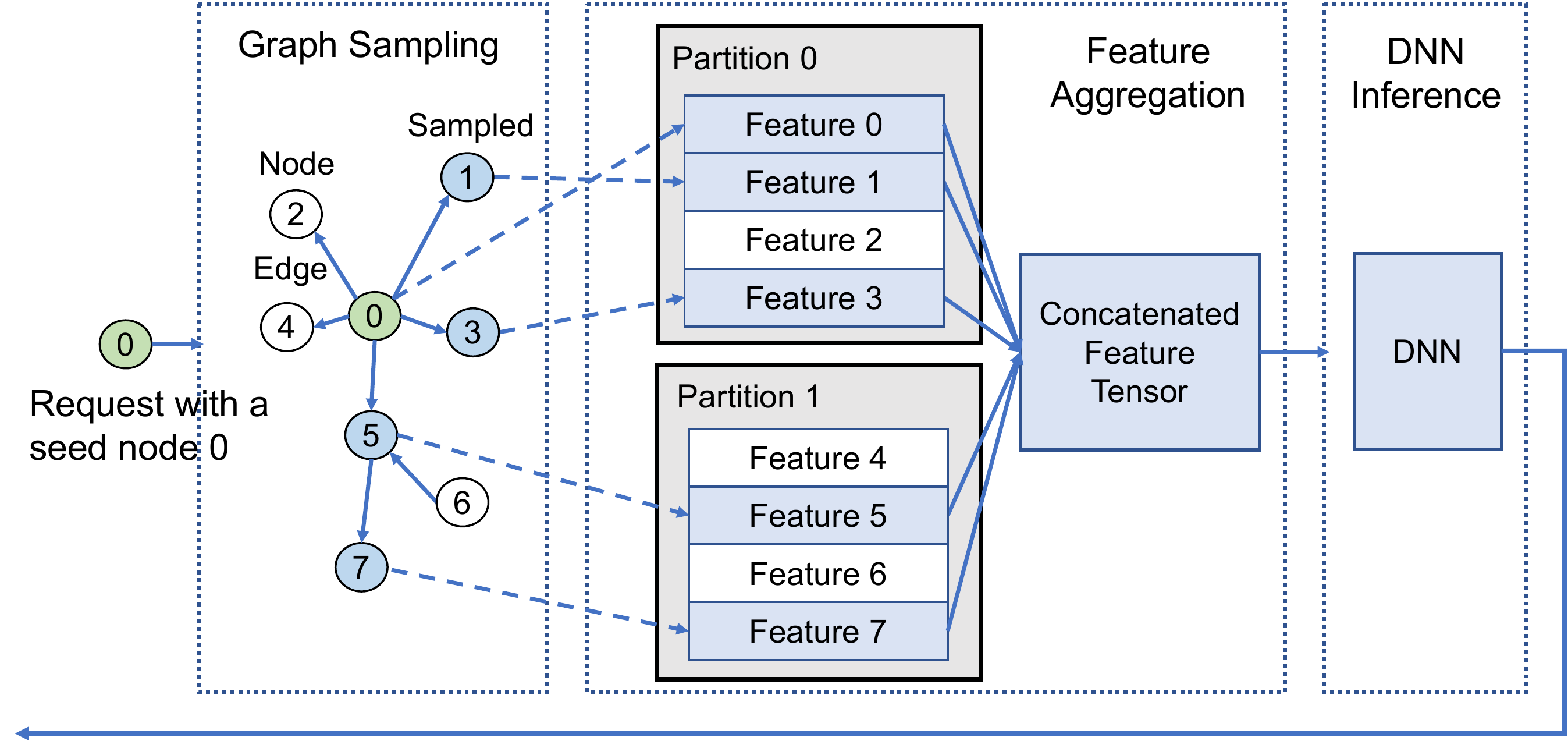}
  \caption{Overview of GNN serving computation}
  \label{fig:background} 
\end{figure}

\F\ref{fig:background} gives shows the GNN serving computation, assuming a 2-layer sampling function. After receiving a GNN request with seed node~0, the system samples its neighbors and returns the layer-1 sampled nodes (1, 3, 5). When it reaches the layer-2 neighbors, it probabilistically samples node~7. After that, it collects the features for all sampled nodes (layer~0, node~0; layer~1, nodes~1, 3, 5; layer~2, node~7), potentially from different devices. It then concatenates the collected features, and the feature tensor is used for DNN inference. Finally, the inference results are returned to the user.

In practice, GNN serving must handle large graphs with many features: \eg MAG240M~\cite{hu2021ogb}, a heterogeneous academic graph dataset, has 240~million graph nodes, 1.7~billion edges, and 768-dimensional feature vectors for each node. Our production internet graph dataset has billions of graph nodes, with feature sizes totalling tens of TBs.

Therefore, features must be partitioned and distributed across servers. Each server comprises of multiple GPUs and dozens of CPU. Devices are inter-connected by a heterogeneous distributed NUMA fabric with NVLink, PCIe, Ethernet, and InfiniBand links~\cite{li2023pond}.  

\subsection{Challenges in large-scale GNN serving}
\label{sec:motivation:challenges}

\begin{figure}[t]
    \centering
    \begin{subfigure}{0.48\linewidth}
        \includegraphics[width=\linewidth]{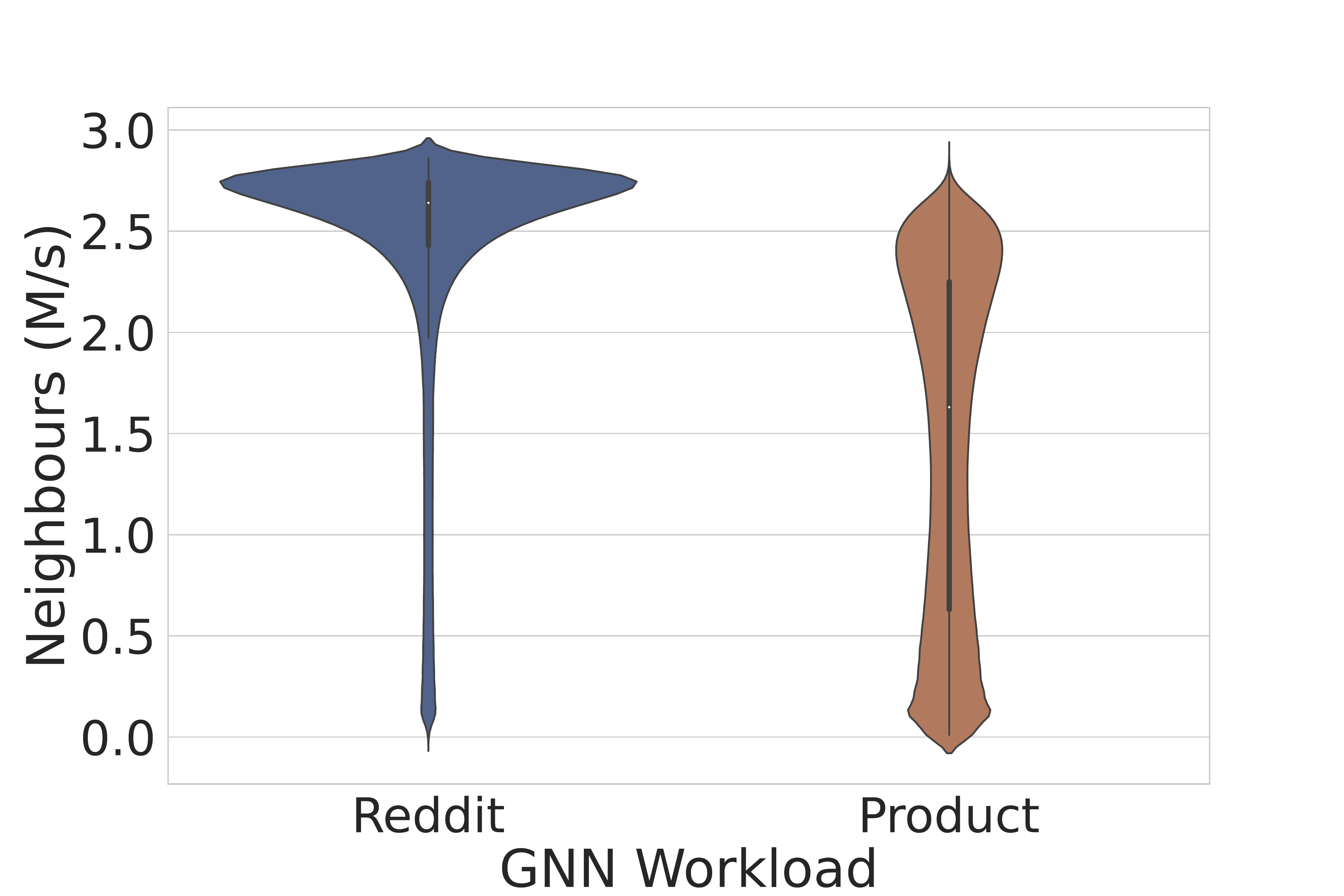}
        \caption{Sampling [25,10] neighbors}
        \label{fig:25-10-neighbours}
    \end{subfigure}
    \begin{subfigure}{0.48\linewidth}
        \includegraphics[width=\linewidth]{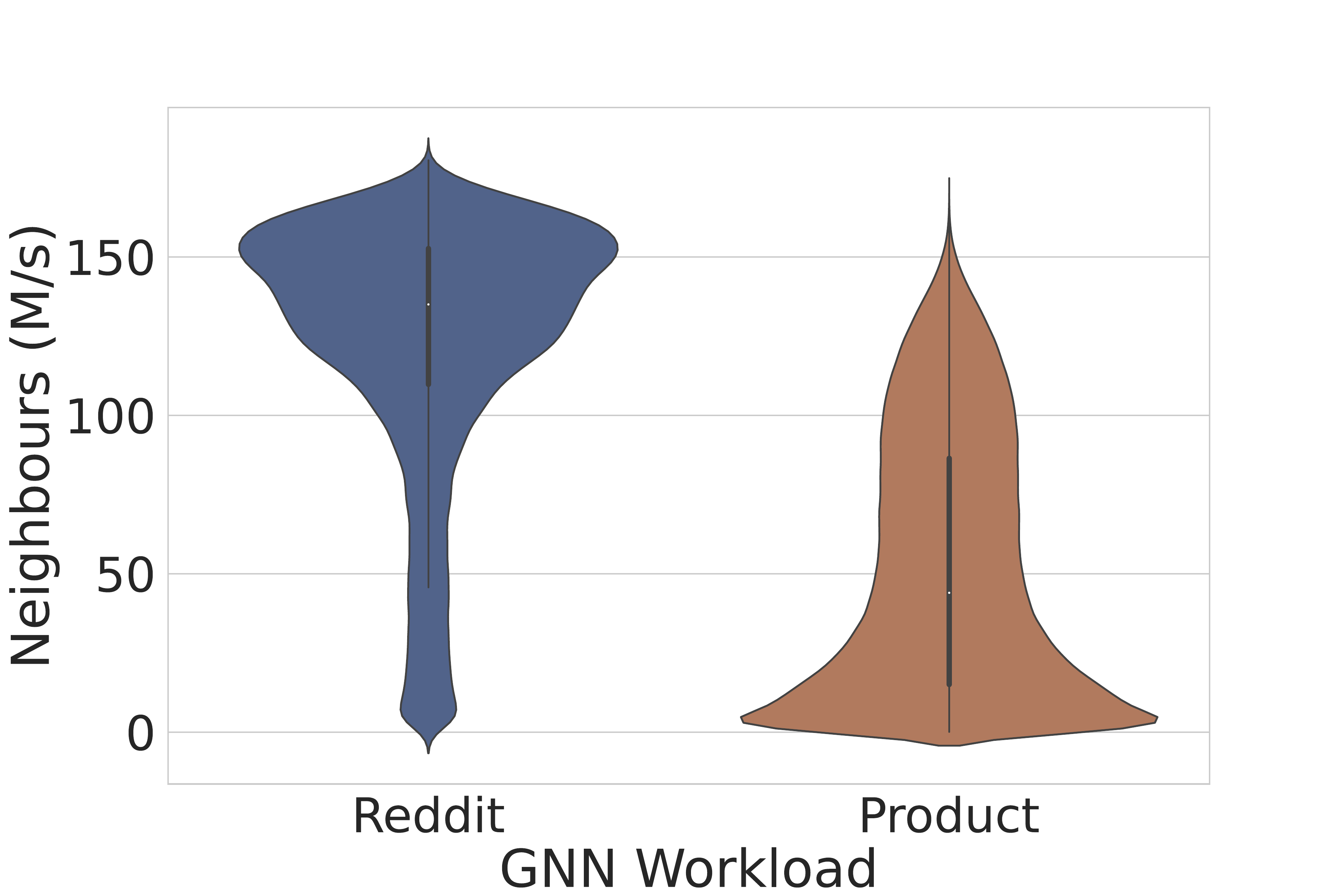}
        \caption{Sampling [50,35] neighbors}
        \label{fig:50-35-neighbours}
    \end{subfigure}
    \caption{Number of sampled neighbors (Reddit and Product) }
\end{figure}

\begin{figure}[t]
    \centering
    \begin{subfigure}{0.48\linewidth}
        \includegraphics[width=\linewidth]{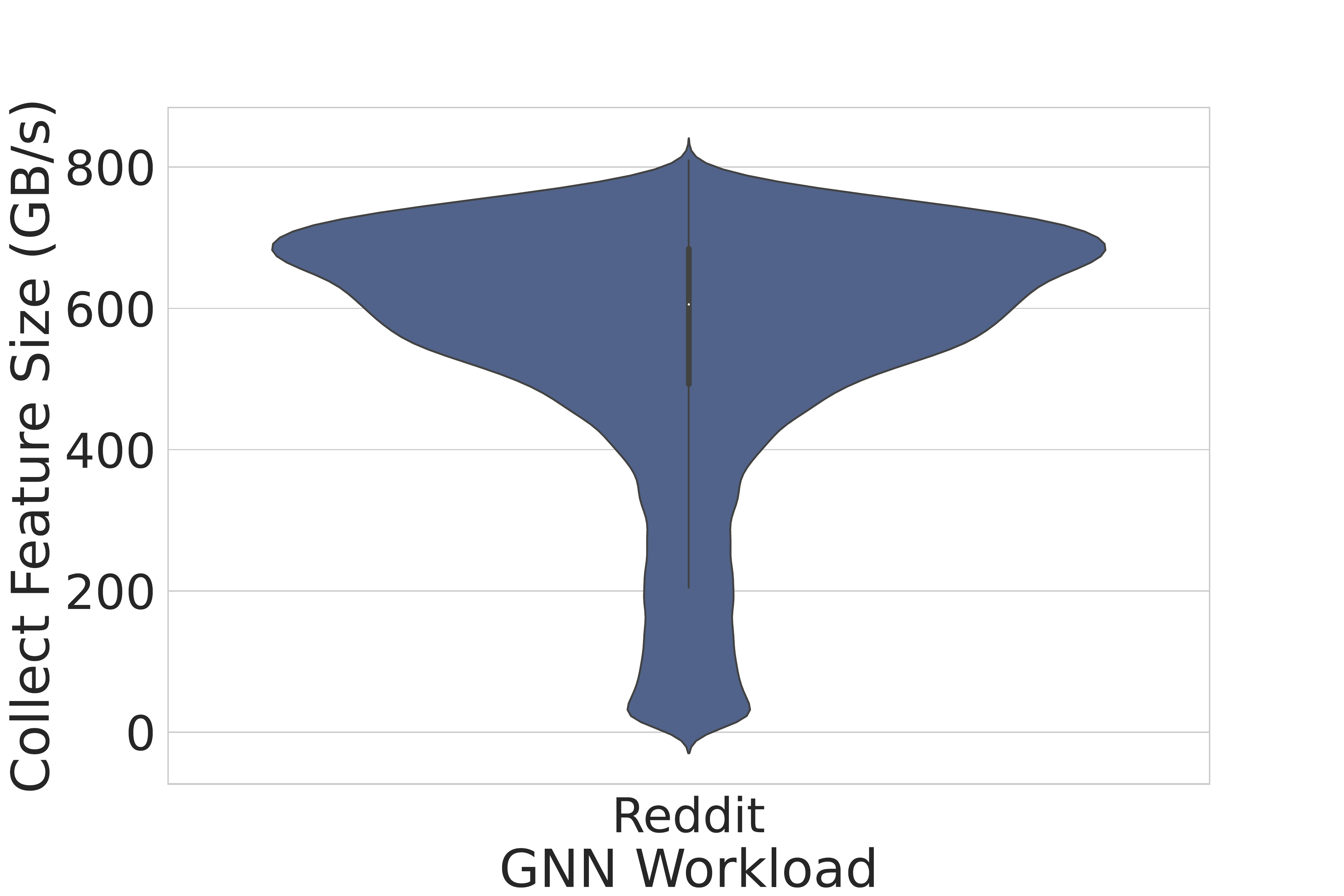}
        \caption{Reddit graph}
        \label{fig:reddit-feature-size}
    \end{subfigure}
    \begin{subfigure}{0.48\linewidth}
        \includegraphics[width=\linewidth]{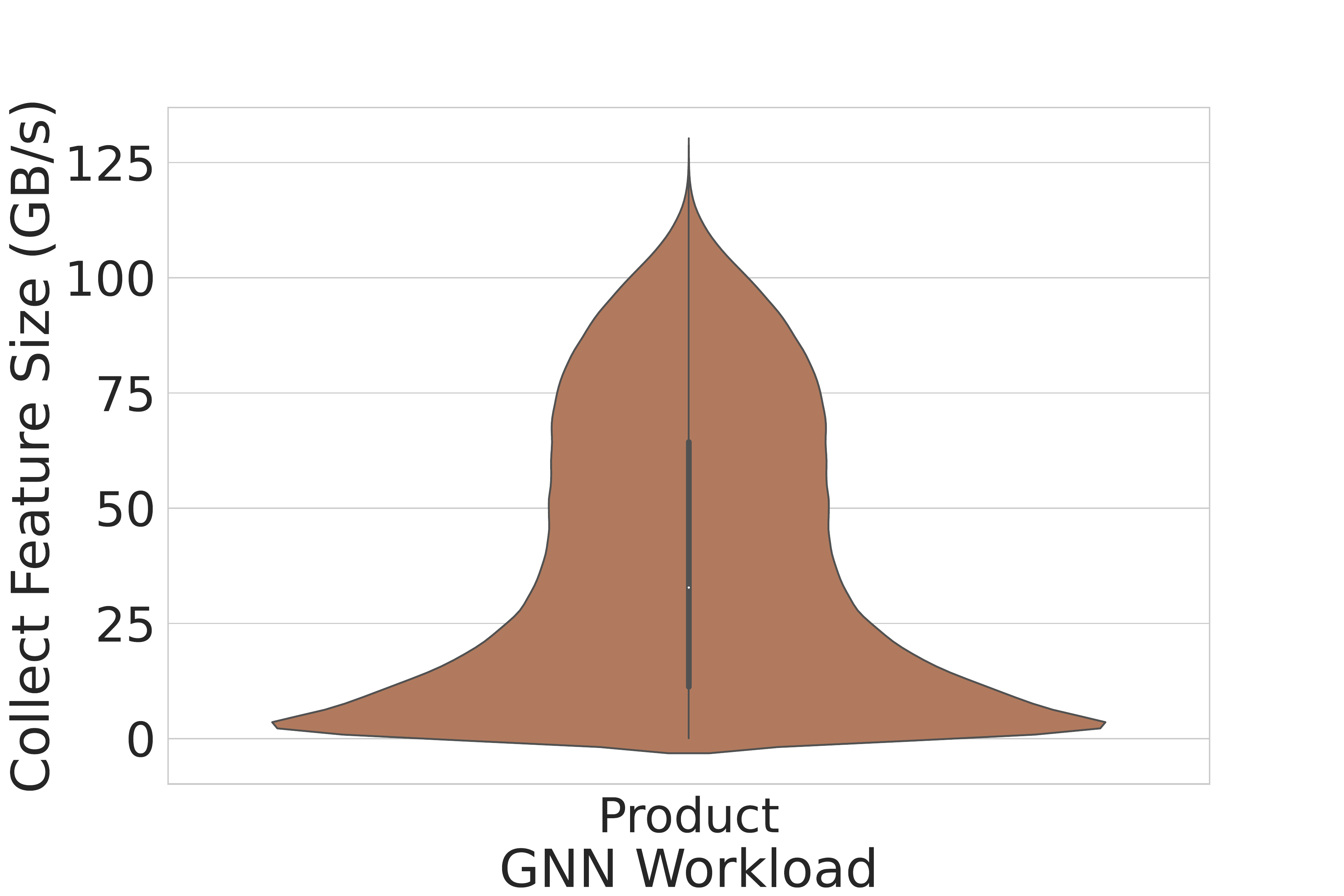}
        \caption{Product graph}
        \label{fig:products-feature-size}
    \end{subfigure}
    \caption{Size of aggregated features}
\end{figure}

Despite the processing scale, GNN requests must be served with low latency. For example, recommender systems  must process thousands of requests within 15 ms~\cite{zhou2018deep}, stream processing handles millions of requests in milliseconds~\cite{cameo2021, mai2018chi}, fraud detection systems must handle millions of requests within 20\unit{ms}~\cite{xu2021towards}, and route planning applications process tens of millions of requests within 100\unit{ms}~\cite{9534153}.

Due to the irregularity of the computation, it is challenging to achieve these latency goals. Real-world graph exhibit a high degree of skew in the number of neighbors associated with the graph nodes. When performing multi-layer neighbor sampling for a batch of GNN requests, systems may process substantially varying numbers of sampled neighbors and aggregated feature sizes.  

We show this variability in sampled neighbors for two real-world graphs when handling a batch of 100,000 GNN requests. Each request requires sampling 25 layer-1 neighbors and 10 layer-2 neighbors. \F\ref{fig:25-10-neighbours} shows that, for the Reddit graph~\cite{hamilton2017inductive}, the number of sampled neighbors ranges from 3,000 to 3,000,000, with the majority falling between 2,000,000 and 2,800,000. For the Product graph~\cite{yang2012defining}, the number of sampled neighbors ranges from 4,000 to 2,600,000. 

This variability makes it challenging to map the sampling computation to a single type of device: CPU-based sampling provides predictably low latency, as CPUs can efficiently access graph data distributed across a large amount of memory, but their limited parallelism reduces throughput; in contrast, GPU-based sampling achieves higher throughput, but only when the GNN request samples many neighbors. A large number of neighbors fully utilizes the high degree of parallelism of GPUs and amortizes their higher start-up and data movement costs.

In addition to graph skew, the number of sampled neighbors is highly sensitive to the graph sampling configuration (\ie the numbers of sampled layers and the number of neighbors per layer). \F\ref{fig:50-35-neighbours} shows that, after adjusting the sampling configuration to include 50 layer-1 neighbors and 35 layer-2 neighbors, the distribution of sampled neighbors changes substantially: for the Reddit graph, the number of sampled neighbors now ranges from 10~million to 175~million; while for the Products graph, it varies between 2~million and 150~million, with the majority around 5~million.

We also examine the variability in the total size of aggregated features. \F\ref{fig:reddit-feature-size} shows that the aggregated feature size for the Reddit graph ranges from 36\unit{GB} to 800\unit{GB}; for the Product graph (\F\ref{fig:products-feature-size}), it ranges from 3\unit{GB} to 110\unit{GB}.

When using GPUs, all these features must be loaded into GPU memory for subsequent DNN computation. Transferring these features over the PCIe bus (with 16--32\unit{GB/s} bandwidth) incurs latencies from hundreds of milliseconds to tens of seconds. Such latencies are significantly higher than the GPU-based DNN computation time (usually in the tens of milliseconds), making feature aggregation a bottleneck.

\subsection{GPU-based GNN serving}
\label{sec:motivation:related_work}



GNN serving systems require predictable low latency processing when exploiting GPUs. Existing GNN systems (\eg PyG~\cite{pyg}, DGL~\cite{dgl}, GNNLab~\cite{gnnlab}, BGL~\cite{bgl}) use GPUs for feature aggregation, and NextDoor~\cite{nextdoor} uses GPUs for accelerating graph sampling. These systems however suffer limitations when using GPUs for serving:

\mypar{Predictable latency on GPUs} Proposals to exploit GPUs for GNN sampling exist. NextDoor~\cite{nextdoor} necessitates a large batch of seed nodes to fully utilize GPUs, but adversely affects latency performance in GNN serving. 

DNN serving systems such as Clipper~\cite{clipper} and Clockwork~\cite{clockwock} use \emph{dynamic batching} to reduce request latencies. They monitor the incoming DNN inference requests and construct dynamically-sized batches that can be processed by a given latency deadline. Such approaches, however, assume a constant computation and communication effort for a single request that targets a given DNN model: for a DNN inference request, the input data (\eg image or text) is of a fixed size and leads to the same amount of activation data. This predictability makes it easier to aggregate requests until a given latency target is reached.

As we have shown in \S\ref{sec:motivation:challenges}, GNN inference requests, however, require varying computational and communication resources. Since batches contain different graph seed nodes, there is a variance in the size of sampled graph nodes and thus aggregated features. This irregular computation makes simple batching techniques that assume a fixed cost per inference request infeasible.


\mypar{Feature assignment to GPUs} Since feature data in GNN serving is large (see \S\ref{sec:motivation:challenges}), the data must be distributed across GPU servers. Existing GNN systems, including DSP\cite{cai2023dsp}, BGL\cite{bgl}, GNNLab\cite{gnnlab}, \emph{cache} popular features in GPUs, which requires a decision on feature popularity: GNNLab\cite{gnnlab} estimates feature popularity by counting the feature's access frequency during model training; BGL\cite{bgl} ranks feature popularity based on their node in/out degrees in the graph.

However, such approaches are ineffective for GNN serving scenarios. When allocating features to servers, GNN serving systems cannot exploit prior information from training: the seed nodes during training are selected deliberately to follow a uniform distribution, which maximizes model accuracy. In contrast, seed nodes in GNN inference requests follow real-world skewed distributions~\cite{kondor2008skew}. Training dataset also only includes a small subset of potential seed nodes (20\%--30\% in the OGB benchmark).


In addition, any method for feature assignment in GNN serving must take multi-layer neighbor sampling into account, otherwise the calculated feature popularity will deviate from those observed when serving GNNs.

\section{Workload Aware GNN Serving}
\label{sec:design}

Next we introduce our idea of workload-aware GNN serving~(\S\ref{sec:design:overview}) and give an overview of \sys's design~(\S\ref{sec:design:architecture}).

\subsection{Overview}
\label{sec:design:overview}


Our analysis in \S\ref{sec:motivation:challenges} reveals that the effectiveness of using GPUs for GNN serving depends on the properties of the graph. Therefore, we want to explore a design for a GNN serving system that is \emph{workload aware}, \ie the system makes decisions regarding the compute and data allocation to GPUs that depend on the graph properties.

Open challenges when realizing this idea is to decide (i)~how and (ii)~when to collect information about the workload. Our approach is to pre-compute \emph{workload metrics} that capture properties of the graph used for GNN serving. If the system pre-computes appropriate metrics at deployment when the graph data is available, it can use the metrics for principled decision-making, both at deployment time when having to partition feature data from the graph across GPU servers and at runtime when assigning GNN inference computation to GPU and CPU devices. The cost of pre-computation of these metrics can be amortized across the execution of GNN requests.

We exploit two workload metrics in \sys's design:

\mypar{Probabilistic sampled sub-graph size~(PSGS)} For a GNN request, the system must predict the computational load of the request to make a decision whether to execute the GNN sampling computation on a GPU or CPU: if the sampled neighborhood yields many nodes, GPU-based sampling is more efficient; if it results in few nodes, the sampling task can be executed by a CPU core with lower latency.

The PSGS metric estimates the number of sampled nodes for a given seed node in a graph, and the system can use it to allocate sampling tasks to the most appropriate device. It can be pre-calculated efficiently by GPUs(see \S\ref{sec:psgs:calculation}). The pre-calculated values are stored in a lookup table, which fits into GPU memory~(see \S\ref{sec:eval:psgs}) and is consulted by the system at runtime.

\mypar{Feature access probability~(FAP)} The bulk of the data movement when serving GNN requests is due to the access of feature data. To prevent feature collection from becoming a communication bottleneck, the system must place features close (in terms of the NUMA/network topology) to the GPUs that access them. If a feature is popular, \ie it has a high probability of access for any given GNN request, it should place within the NUMA/network topology in such a way that allows for lower latency access.

The FAP metric estimates the access likelihood of any feature data in the graph. It is  calculated by GPUs by implementing graph sampling as sparse matrix multiplication~(see \S\ref{sec:fap}). Based on the FAP value, the system can place feature data across multiple levels of the NUMA/network topology (from lowest to highest latency access): (1)~local GPU; (2)~GPU in the same server, interconnected via NVLink~\cite{he2022scgraph}; (3)~GPU in the same server, interconnect via PCIe; (4)~GPU in the different server, interconnect with InfiniBand.

\begin{figure}[t]
  \centering
  \includegraphics[width=\linewidth]{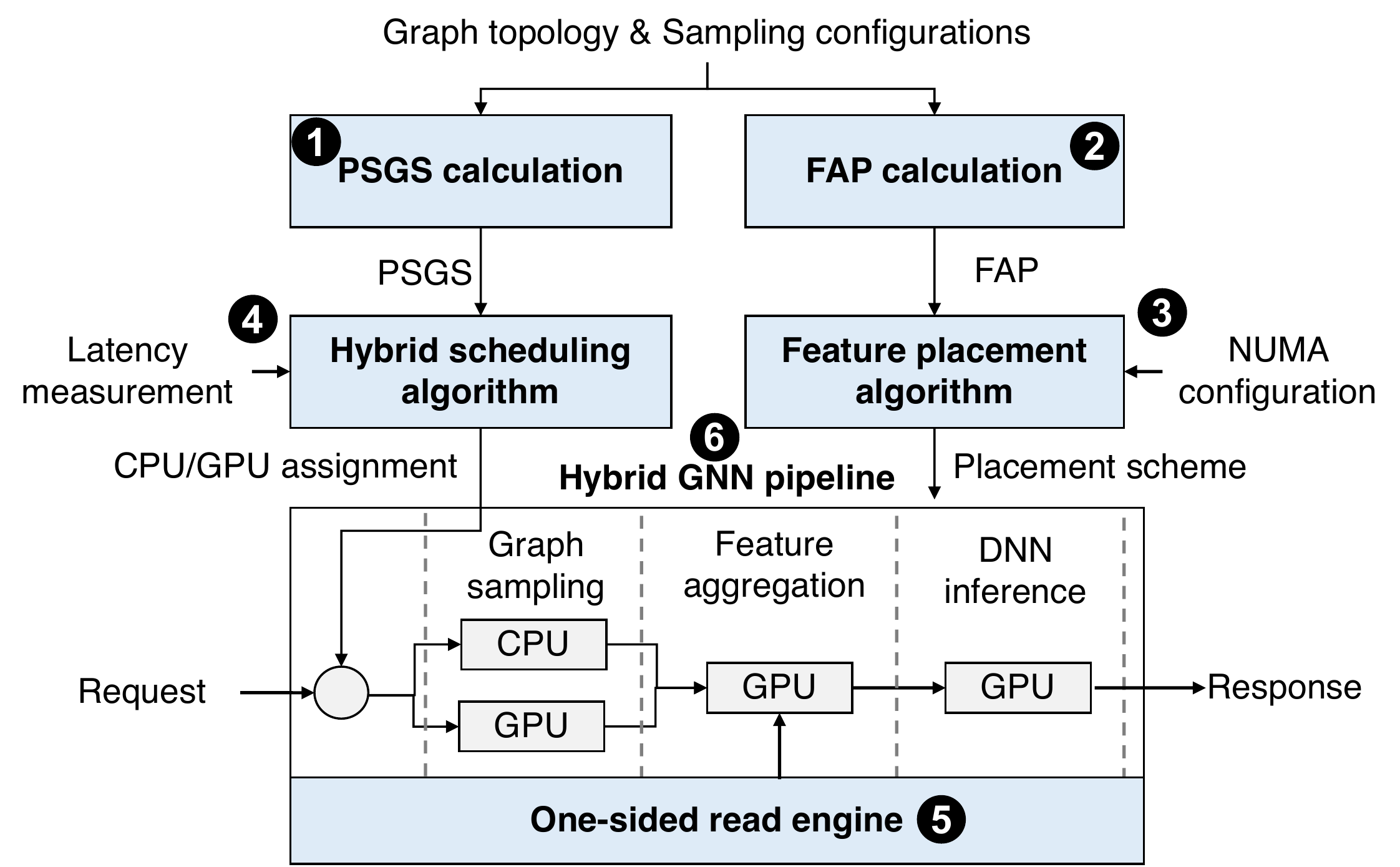}
  \caption{\sys design}\label{fig:overview} 
\end{figure}

\subsection{Design}
\label{sec:design:architecture}

Next describe the design of \sys, a distributed GNN serving system for GPUs that uses the PSGS and FAP metrics for workload awareness. Figure~\ref{fig:overview} shows the design: \sys takes a \emph{graph topology} and \emph{sampling configurations} as input at deployment time. These are used to pre-calculate the PSGS~\myc{1} and FAP metrics~\myc{2}, which is done efficiently by parallelizing the computation using GPUs (\S\ref{sec:psgs:calculation}). For the PSGS metric, \sys analyzes the relationship between the PSGS value and latency measurement of different emulation batches using a serving workload generator. It generates the relationship between PSGS and latency measurement, which allows \sys to choose PSGS that can guide the assignment of GNN sampling to GPUs and CPUs.

After the FAP metric is calculated, a \emph{feature placement algorithm}~\myc{3} uses it, in combination with information about the NUMA/network topology of the deployment, as input to decide on the feature assignment. It sorts the features based on the FAP metric and partitions and replicates features across the topology: it partitions popular features among GPUs, connected through NVLink and InfiniBand, thus caching popular features in GPUs and reducing PCIe traffic; for GPUs without NVLink and InfiniBand, \sys replicates popular features to increase locality of access.

When processing GNN inference requests, a \emph{hybrid scheduling algorithm}~\myc{4} dynamically assigns graph sampling tasks to GPUs and CPUs. It performs a PSGS lookup for each GNN request, and only assigns the request to GPUs when it improves throughput without increasing latency.

GNN requests are processed by a \emph{hybrid GNN pipeline}~\myc{5}, which efficiently exploit a large number of GPU and CPU cores in executing different GNN computation stages (\ie graph sampling, feature aggregation, and DNN inference), achieving high-throughput GNN request process.

As part of the pipeline, the features needed to execute feature aggregation tasks are collected by a \emph{one-sided read engine} \myc{6}. If a feature can be accessed via NVLink and InfiniBand, the engine directly reads the feature from a peer GPUs/CPUs, avoiding interrupting CPUs and minimizing memory copies.

\section{Workload-Aware GNN Sampling}
\label{sec:psgs}

In this section, we introduce the computation of PSGS (Section \ref{sec:psgs:calculation}), describe how PSGS contributes to GNN sampling in achieving consistent latency performance (Section \ref{sec:hybrid_sampling}), and discuss how GNN serving pipeline can achieve high throughput (Section \ref{sec:hybrid_pipeline}).

\subsection{Estimation of probabilistic sampled subgraph size}
\label{sec:psgs:calculation}


\begin{figure}[t]
    \centering
    \includegraphics[width=\linewidth]{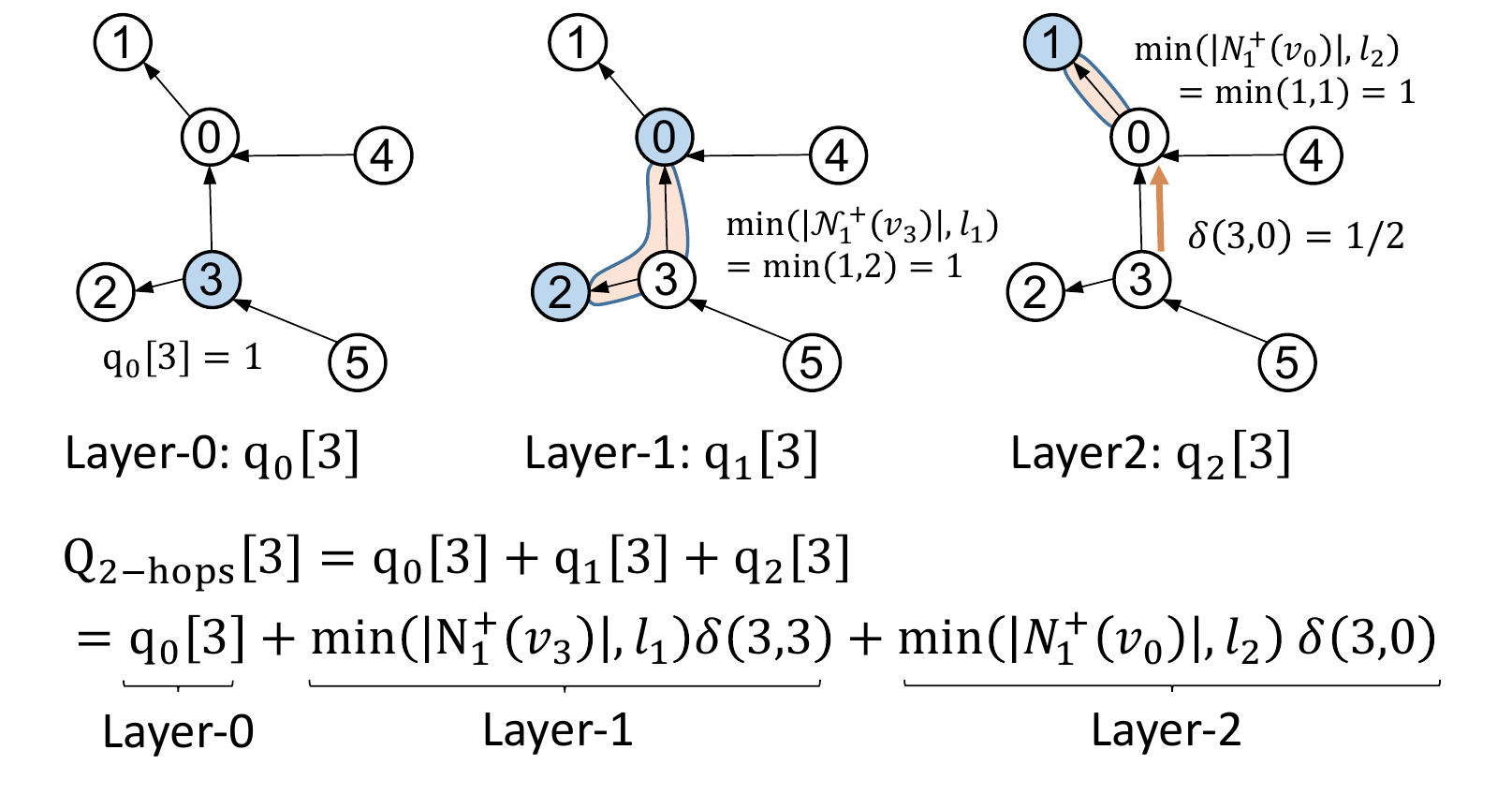}
    \caption{Probabilistic Sampled Sub-graph Size of Node 3}
    \label{fig:p-sgs}
\end{figure}

The estimation of the PSGS must account for the configuration of a probabilistic multi-layer neighbor sampling method. In the following, we use an example to describe how this configuration is involved in computing the PSGS and then give a formal definition.

\mypar{Example} Fig~\ref{fig:p-sgs} shows an example of the calculation of the PSGS of node 3 ($Q_{2-hops[3]}$). Assume the maximum sample size of hop-1 and hop-2 are 2 and 1 respectively. $Q_{2-hops}[3]$ is the sum of $\text{q}_0[3]$, $\text{q}_1[3]$ and $\text{q}_2[3]$, which represents the expected subgraph size at hop-0, hop-1 and hop-2 respectively. 

$\text{q}_0[3]$ is always the subgraph with only the seed node, so the size is 1. The size of the hop-1 subgraph is 1, which is the minimum of the hop-1 neighbourhood size of 2 and the maximum sample size of 1, so $\text{q}_1[3]$ is 1. The probability that transits from node 3 to node 0 is $1/2$ and the subgraph size from 0 is 1, so $\text{q}_2[3]$ is $1 \times 1/2 = 1/2$.

\mypar{Construction algorithm} Specifically, the PSGS in $k$-hop sampling for a node $i$, denoted as ${Q}_{K-\text{hop}}[i]$, is defined as:
$
    \text{Q}_{K-\text{hops}}[i]=\sum_{k=0}^{K}\text{q}_k[i]
$
, where

\begin{equation*}
\label{prob_eq}
\text{q}_k[i] = \begin{cases}
1, & k = 0
 \\
 \sum_{v_j\in{N}^{+}_{k-1}(v_i)}min(|{N}^{+}_{1}(v_j)|,l_k)\delta_{k-1}(i, j)\
, & k > 0
\end{cases}
\end{equation*}



$\text{q}_0[i]$ refers to the probability sampled sub-graph size(PSGS) that each point can sample at the 0\textsuperscript{th}-layer, which is essentially the point itself. Therefore, $\text{q}_0[i]=1$. $\text{q}_k[i]$ represents the PSGS of node $i$ at the k\textsuperscript{th}-hop. 

${N}^{+}_{k-1}(v_i)$ defines the set of the k\textsuperscript{th}-hop out-neighbors of node $i$, which is the set of all nodes that can be sampled from node $i$ in the k\textsuperscript{th}-hop. $\delta_{k}(i, j)$ is the probability of sampling $v_j$ from $v_i$ at the k\textsuperscript{th}-hop (\ie the transition probability). Both ${N}^{+}_{k-1}(v_i)$ and $\delta_{k}(i, j)$ can be obtained by calculating the k\textsuperscript{th}-order weighted adjacency matrix $A^k=\prod^{k}A$. ${N}^{+}_{k-1}(v_i)$ is the set of column indices corresponding to all non-zero elements in the $i$-th row of matrix $A^k$, and $\delta_{k}(i, j)=A^k[i][j]$

The output of this algorithm, $\text{Q}_{K-\text{hops}}$, is a lookup table stored in memory as an array, with a space complexity of $O(|V|)$. The time complexity for querying is $O(1)$.

\mypar{Computation cost} When analyzing sampled sub-graph size, \sys must compute the PSGS metric for each graph node. For the entire graph, the dominating computation cost lies in finding the set of K-hop out-degree neighbors of each node and the transition probabilities between each node in the graph at the K\textsuperscript{th}-hop. This requires calculating the K\textsuperscript{th} order weighted adjacency matrix $A^K$. The time complexity of this calculation using a CPU for serial matrix multiplication is $O(k|V|^3)$. 

Since most real-world graphs are sparse (\eg the adjacency matrix is sparse)~cite{evidence}, \sys implements this process using a GPU. It employs CUDA's sparse matrix multiplication operator, which reduces the time complexity to $O(k|V||E|)$, where $|E| << |V|^2$ in sparse matrices. When analyzing a graph with hundreds of millions of graph nodes, the PSGS computation on a GPU can finish within minutes. 

\subsection{PSGS-guided hybrid sampling}
\label{sec:hybrid_sampling}

\sys can support GPUs to achieve predictable latency performance with the aid of PSGS. The main idea is to analyze the relationship between the PSGS and request processing latency offline. After being deployed, \sys can monitor the seed nodes of GNN requests and estimate the PSGS of these requests. With the estimated PSGS, \sys predicts the latency required for processing the requests. It assigns the request processing to GPUs only when it enhances throughput with predictable low latency.

\subsubsection{PSGS and processing latency}

\begin{figure}[t]
  \includegraphics[width=0.48\textwidth]{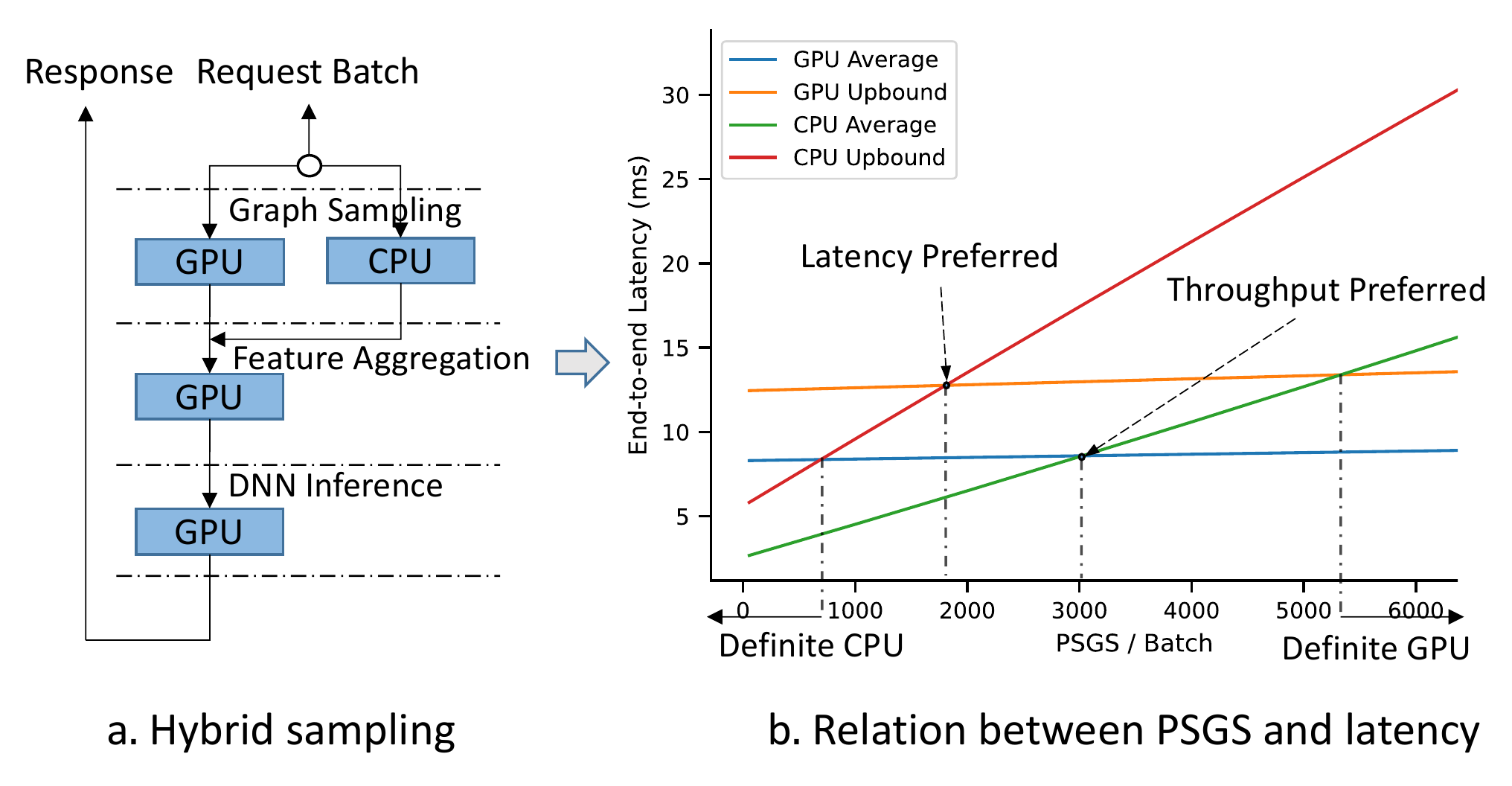}
  \caption{Relation between PSGS and GNN serving latency}
  \label{fig:sgs-crosspoint} 
\end{figure}

We aim to predict the relationship between PSGS and request processing latency. To do this, we generate multiple GNN request batches with varying PSGS values. We then measure the processing latency of these batches in a hybrid sampling pipeline, shown by Figure \ref{fig:sgs-crosspoint}(a). In this pipeline, graph sampling can be assigned to either CPUs or GPUs, while feature aggregation and DNN inference are assigned exclusively to GPUs.

\sys is designed to ensure that offline latency measurements are accurate and consistent with those in a serving scenario. To achieve this, \sys incorporates a serving workload generator that conducts latency measurements when both CPUs and GPUs are near full utilization, with no queuing in the pipeline. The serving workload generator continuously produces batches until there are a sufficient number of latency measurements for each PSGS, thus ensuring the reliability of the measurements.

After gathering an adequate number of latency measurements, \sys generates a figure that illustrates the relationship between PSGS and the end-to-end processing latency of the hybrid sampling pipeline, as demonstrated in Figure \ref{fig:sgs-crosspoint}(b). In this figure, we visualize both the average latency and the maximum latency achieved when using either GPUs or CPUs for GNN sampling. The maximum latency measurement enables \sys to evaluate how to select a PSGS that complies with a \emph{latency bound}, while the average latency measurement allows \sys to choose a PSGS that targets a specific \emph{latency goal}.
\newcommand{\myparinew}[1]{\noindent\emph{#1.}\xspace}

In the figure mentioned above, we observe the latency measurement lines intersect at 4 points:
\myparinew{(a)~CPU preferred} Point \myc{1} is where the CPU maximal latency intersects the GPU average latency. For any GNN request with a PSGS smaller than the CPU preferred point, this request can be completed faster on CPUs, even in the worst-case scenario. 
\myparinew{(b)~GPU preferred} Point \myc{2} is where the CPU average latency intersects the GPU maximal latency. For any request with a PSGS larger than this point, sampling can be completed on GPUs with enhanced latency and throughput performance.
\myparinew{(c)~Latency preferred} Point \myc{3} is where the CPU maximal latency line intersects the GPU maximal latency line. If users prioritize bounding latency performance, they can select this cross point to guide the hybrid sampling: any GNN request with a PSGS smaller than the latency preferred point is assigned to CPUs. If larger, it is assigned to GPUs.
\myparinew{(d)~Throughput preferred} Point \myc{4} is where the CPU average latency line intersects the GPU average latency line. If users prioritize increasing throughput, they can choose this cross point to guide the hybrid sampling process.

\subsubsection{GNN serving with PSGS} In the following, we explain how to utilize the selected PSGS value to enable efficient GNN serving while maintaining predictable performance. During GNN serving, the \sys system continuously batches incoming GNN requests, completing the process once a batching deadline is reached. The \sys system then iterates through all seed nodes within this batch, accumulating their PSGS estimations. If the accumulated sum is less than the user's chosen PSGS value, the batch is assigned to CPUs for GNN sampling completion; otherwise, it is assigned to GPUs. This approach ensures that GPUs can deliver predictable low latency, while simultaneously directing the majority of the graph sampling workload to GPUs, thereby increasing throughput.

\subsection{High-throughput hybrid pipelines}
\label{sec:hybrid_pipeline}

Designing high-throughput hybrid pipelines for GNN serving introduces several challenges. In the following, we discuss our design choices that address them:

\mypar{(1)~Multiplexing GNN pipelines in a processor} The processing of GNN requests require both compute-intensive stages (\eg for graph sampling and DNN inference) and communication-intensive stages (\eg for feature aggrgegation). A GNN pipeline can be thus interrupted for communication, leaving the processor idle. To address this, \sys multiplexes multiple pipelines in one processor (\eg with each pipeline running in a CUDA stream). Such a design allows the processor to process multiple requests concurrently, overlapping their computation and communication tasks~\cite{koliousis2019crossbow}.

\mypar{(2)~Sharing the queue for GNN pipelines in a processor} GNN requests with irregular computation patterns lead to diverse processing times on GPUs. To avoid dispatching batches to a slow pipeline, incurring significant queuing delays, \sys creates a queue shared by the pipelines on the same processor. These pipelines compete for requests in the shared queue, avoiding queuing delays and stragglers.

\mypar{(3)~Sharing the graph for GPU pipelines in a server} GNN requests sample large graphs, which consumes substantial memory (\eg 100s of GBs). GPUs, however, have limited memory (typically 16\unit{GB}--80\unit{GB}). To address this, \sys replicates the graph topology in each server and makes all the GPU pipelines share this graph.

To make graph sharing efficient, we implement the shared graph using the GPU's unified virtual addressing~(UVA) memory. Each graph partition is implemented as a pinned memory block and directly mapped to the GPU's memory space.

\section{Workload-aware Feature Placement}

In this section, we describe how \sys computes the feature access probability~(\S\ref{sec:fap}), places features on GPU servers~(\S\ref{sec:placement}), and uses efficient one-sided GPU reads to access features~(\S\ref{sec:dma}).


\subsection{Estimation of feature access probabilities}
\label{sec:fap}

\begin{figure}[t]
    \centering
    \includegraphics[width=\linewidth]{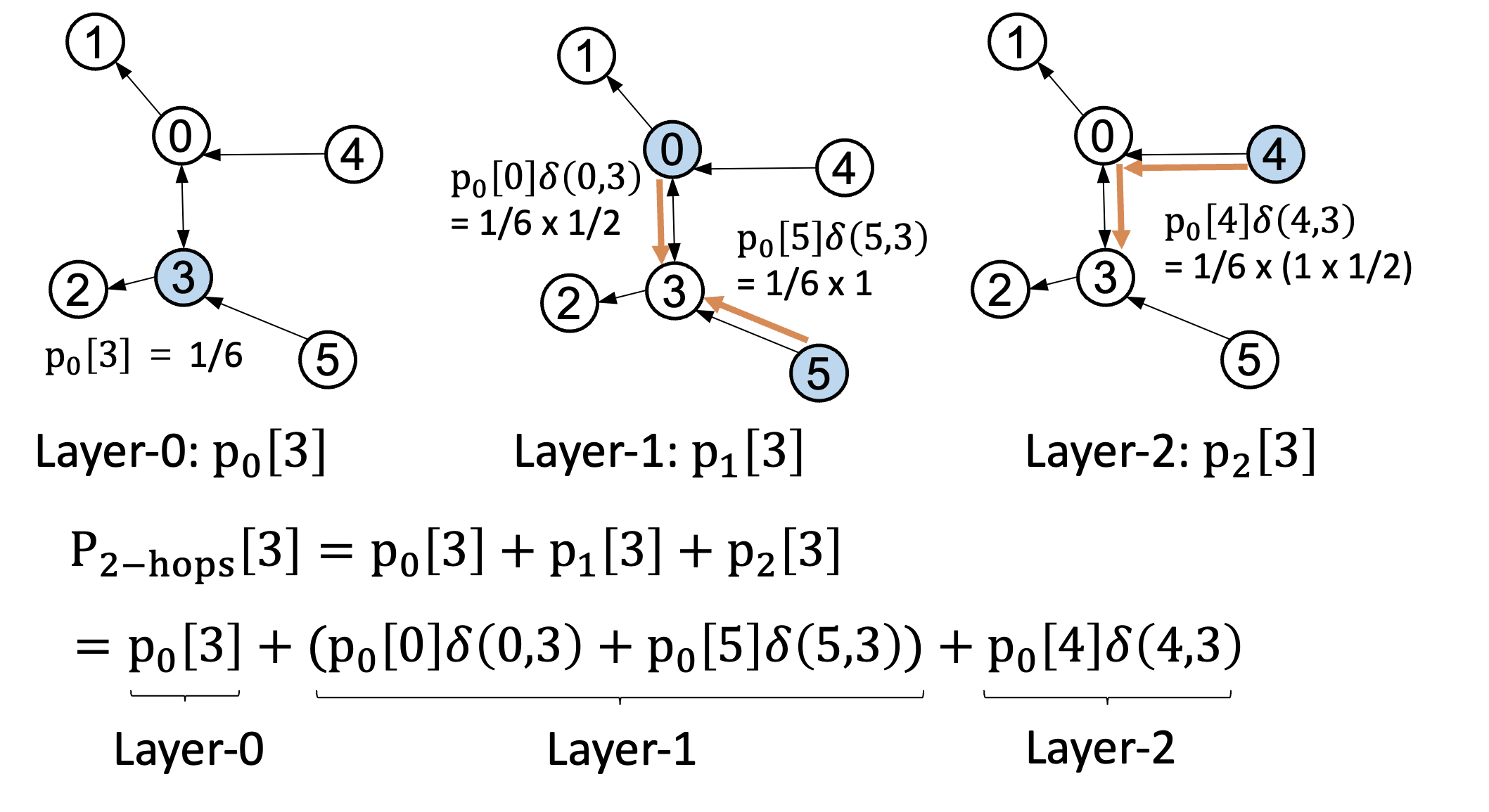}
    \caption{Computing the feature access probability for node~3}
    \label{fig:fap}
\end{figure}

The estimation of the feature access probability~(FAP) is based on the following observation: a node's feature is fetched from memory when the node is in the k-hop sampling subgraph of input seed nodes. Consequently, the more subgraphs a node feature is involved in, the higher the probability that the node feature is accessed. In the following, we use an example to explain the computation of this probability and then present a formal definition.

\mypar{Example} Consider node~3 in the directed graph with equal edge weights shown in \F\ref{fig:fap}. We want to compute the probability of node~3 being sampled as a neighbor within 2 hops from other nodes, denoted as $\text{p}_{2-\text{hops}}[3]$. It is the sum of $p_0[3]$, $\text{p}_1[3]$, and $\text{p}_2[3]$, which represent the probabilities that node~3 is sampled within the 0\textsuperscript{th}, 1\textsuperscript{st}, and 2\textsuperscript{nd} layers, respectively.

Specifically, $\text{p}_0[3] = \frac{1}{6}$ is the probability that node~3 is selected from the 6 nodes as a seed node; $\text{p}_1[3]$ is the sum of the probabilities that node 3 is sampled from its one-hop neighbors (nodes~0 and~3). The probability that node~3 is sampled from node~0 is $\frac{1}{6} \times \frac{1}{2}$, and the probability from node~3 is $\frac{1}{6} \times 1$; and $\text{p}_2[3]$ is the probability that node~3 is sampled from its two-hop neighbor (node~4) via its one-hop neighbor (node~0). The probability that node~4 is sampled at the 0\textsuperscript{th}-layer is $\frac{1}{6}$, and the probabilities of transitioning from node~4 to node~0 and from node~0 to node~3 are $1$ and $\frac{1}{2}$, respectively. Thus, $\text{p}_2[3]$ is $\frac{1}{6} \times 1 \times \frac{1}{2} = \frac{1}{12}$.

\mypar{FAP definition and computation} Generally, the FAP of a node~$v$ sampled within K-hops neighbor 
is computed recursively as follows:
$
\label{eq:fap_prob}
    \text{P}_{K-\text{hops}}[i]=\sum_{k=0}^{K}\text{p}_k[i]
$
, where

\begin{equation*}
\text{p}_k[i] = \begin{cases}
c, & k = 0
 \\
 \sum_{v_j\in{N}^{-}_{k}(v_i)}\text{p}_{0}(j)\
\delta_k(j, i), & k > 0
\end{cases}
\end{equation*}

$
$

\noindent
$\text{p}_0[i]$ is the probability that node~$i$ is sampled at the 0\textsuperscript{th}-layer and $\sum^{|V|}_{i=0}\text{q}_0[i]=1$, \ie that node~$i$ is directly requested as a seed node. If the probability of each node becoming a seed node is equal, then $\text{p}_0[i]=\frac{1}{|V|}$. Users can also set $\text{p}_0[i]$ based on the actual dataset; $\text{p}_k[i]$ denotes the probability that node~$i$ can be sampled from other nodes in the k\textsuperscript{th}-hop. 

${N}^{-}_{k}(v_i)$ defines the set of k\textsuperscript{th}-layer in-neighbors of node~$i$, which is the set of all nodes that can reach node~$i$ in the k\textsuperscript{th}-hop. ${N}^{-}_{k}(v_i)$ can be obtained by calculating the k\textsuperscript{th}-order weighted adjacency matrix $A^k=\prod^{k}A$. ${N}^{-}_{k}(v_i)$ is the set of row indices corresponding to all non-zero elements in the $i$-th column of matrix $A^k$. This requires calculating the K\textsuperscript{th} order weighted adjacency matrix $A^K$. Using the previous analysis from \S\ref{sec:psgs}, the time complexity of this calculation can be optemize by CUDA's sparse matrix multiplication operator to $O(k|V||E|)$.
%
%





\subsection{Feature placement}
\label{sec:placement}

\begin{figure}[t]
    \centering
    \includegraphics[width=\linewidth]{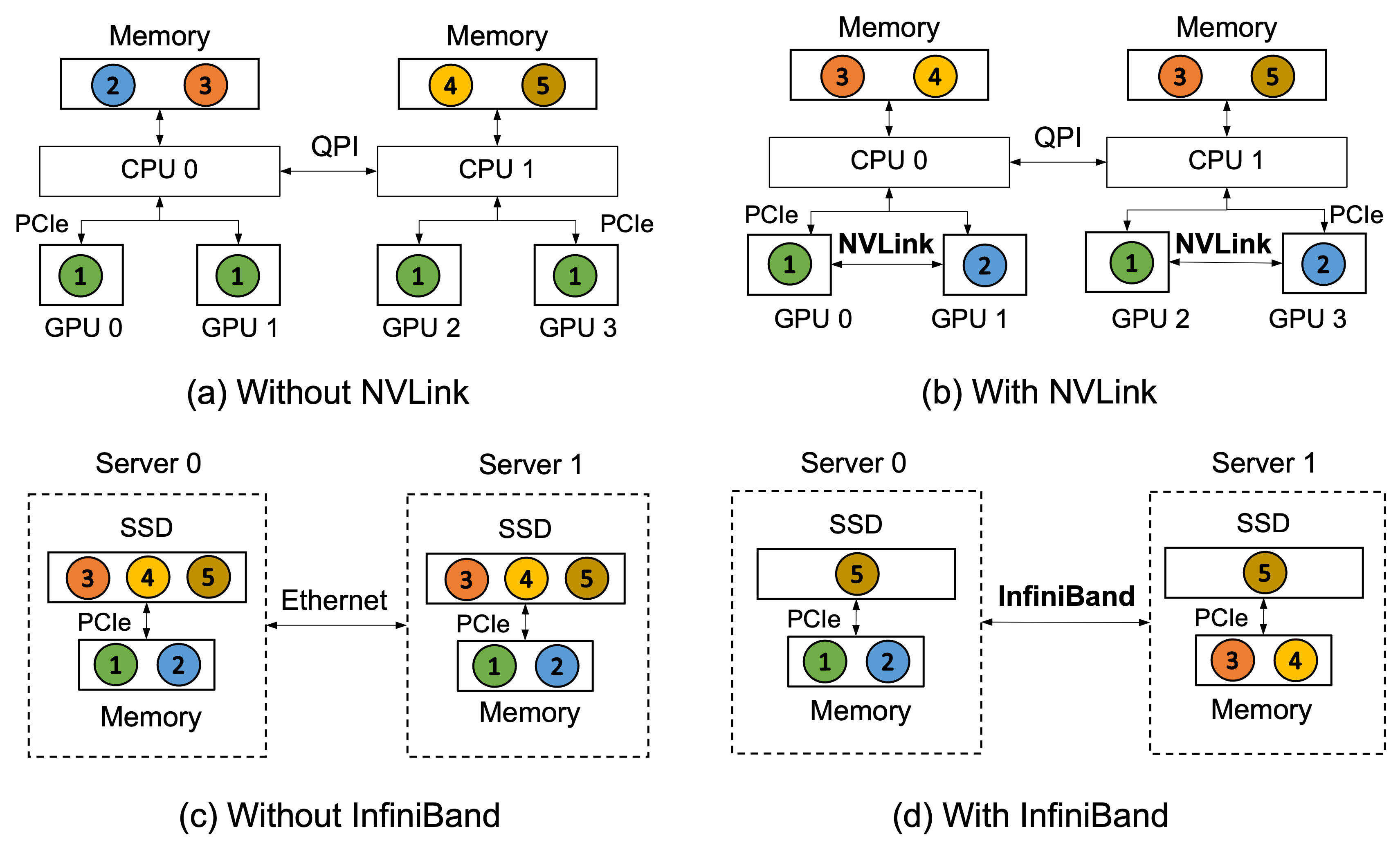}
    \caption{Feature placement scenarios}
    \label{fig:partition}
\end{figure}

\sys uses the FAP metric to place popular features strategically on GPUs. A primary objective of feature placement is to enable GPUs to take advantage of low-latency connectivity, such as NVLink and InfiniBand, to their peer GPUs. This allows GPUs to achieve low-latency access to features when aggregating features.

Minimizing the latency of feature aggregation presents a unique challenge: the features data is large and must be partitioned across servers. The \emph{feature aggregation latency} is determined when \emph{all} sampled features by a request become available in the GPU, allowing it to initiate DNN inference. In other words, this latency is equivalent to the tail latency of the last feature becoming available. This latency-driven minimization target makes the feature placement problem different from GNN training, which instead focuses on cache hit ratios (\eg GNNLab~\cite{gnnlab}, AliGraph~\cite{aligraph} and BGL~\cite{bgl}).

\mypar{Impact of connectivity}  Next, we derive insights from examples that show how NVLink and InfiniBand connectivity impact feature placement. 

\mypari{(a)~Without NVLink} \F\ref{fig:partition}(a) shows feature placement in a server without NVLink. There are 5~features, and their FAP metrics decrease with their ID (\ie feature~0 has the highest FAP value; feature~5 has the lowest). We assume that a server has two NUMA nodes, each with 1~CPU and 2~GPUs. The NUMA nodes are connected using a fast processor interconnect (\eg UPI), and the CPU and GPU are connected using PCIe. The GPU's high-bandwidth memory~(HBM) can hold one feature, and the CPU memory can hold two features.

In this scenario, feature placement is not NVLink aware, and optimizes for data locality only. Consequently, it replicates feature~1 on all GPUs and evenly partitions the remaining features on the CPUs. Consider a GNN request that needs to aggregate features~1 and~2: the feature aggregation latency is determined by the latency of fetching feature~2 from the CPU to the GPU over the PCIe.

\mypari{(b)~With NVLink} \F\ref{fig:partition}(b) shows an improved feature placement that exploits NVLink. As NVLink offers high bandwidth and low data transfers to GPUs within the same NUMA node, fetching a feature over NVLink can be up to 50$\times$ faster than over PCIe. With this in mind, instead of replicating the most popular features on all GPUs, we can partition popular features and assign them to GPUs evenly. For example, feature~1 is placed in GPU~0 and feature~2 is placed in GPU~1. Since accessing data across NUMA nodes is costly, we can replicate features~1 and~2 in the GPUs in both NUMA nodes, still optimizing for data locality.
This optimized feature placement strikes a balance between replication and partitioning, yielding improved feature aggregation latency. Consider again the request that must aggregate features~1 and~2: now GPU 0 fetches feature~2 from its peer GPU 1 over NVLink, while GPU~1 fetches feature~1 from GPU~0 over NVLink. This avoids fetching feature~2 over the slower PCIe bus, reducing aggregation latency.

\mypari{(c)~Without InfiniBand} \F\ref{fig:partition}(c) shows a scenario in which features must be placed across servers. Existing distributed feature placement methods (\eg GNNLab, AliGraph, and BGL) assume that cross-server communication is slow (usually provided by Ethernet). They optimize for data locality, replicating popular features~1 and~2 on both servers and leaving the remaining features in the local disk.

Consider a GNN request that aggregates features~1, 2, and 3: feature~3 must be fetched from disk, incurring slow I/O operations, which increase feature aggregation latency.

\mypari{(d)~With InfiniBand} By making the placement InfiniBand aware, we can trade data locality for a fast InfiniBand link, thus \emph{partitioning} popular features instead of replicating them. We assign features~1 and~2 to server~0 and the other popular features~3 and~4 to server~1.

When executing a GNN request that aggregates features~1, 2, and 3 on a GPU, the GPU can take advantage of the InfiniBand link by fetching feature~3 from the peer server. InfiniBand offers a bandwidth of up to 800\unit{Gbps}, which is 80$\times$ faster than conventional 10-Gbps Ethernet and SSDs. Consequently, feature aggregation latency is substantially improved compared to caching features locally.

\mypar{Placement algorithm} We design an algorithm that takes into account NVLink/InfiniBand connectivity when placing features, minimizing feature aggregation latency. Its key steps are as follows:
(i) \textbf{Sort features:} the placement algorithm begins by sorting all features based on their FAP values. The features have IDs in the range of $0$ to $N$;
(ii) \textbf{Analyze feature capacity per GPU:} the algorithm considers number of features that can be placed in a GPU (denoted as the feature capacity). For this, \sys requires the user to provide the number of GPUs~$G$ in a server, the number of features that can be placed in a GPU~$N_g$, and the number of NUMA nodes~$C$ per server (We only consider the case in which GPUs are connected via NVLink in a NUMA node.) The resulting feature capacity is $\frac{G}{C}N_g$;
(iii) \textbf{Analyze feature capacity per server:} the algorithm analyzes the feature capacity per server, denoted as $N_s$. If InfiniBand is used, $N_s=\frac{G}{C}N_g+N_m$, where $N_m$ represents the number of features that can be placed in server memory; otherwise, $N_s=\frac{G}{C}N_g+N_m+N_d$, where $N_d$ denotes the number of features that can be placed on disk;
(iv) \textbf{Partition and replicate (inter-server):} based on $N_s$ and the number of servers~$S$, the algorithm partitions the most popular features, with each partition containing $N_s$~features. It returns the most popular partition and replicates the features with IDs in the range of $[0:S \times N_s]$ in each server. Finally, it partitions the features with IDs in the range of $(S \times N_s : N]$: for each partitions, it sorts them according to their FAP values and initially places the features in the server memory. After exhausting memory, it places the remaining features on disk; and (v) \textbf{Partition and replicate (intra-server):} for each server, the algorithm replicates the features in the range of $[0:\frac{G}{C}N_g]$ across NUMA nodes. Within each NUMA node, it partitions the features, evenly assigning them to GPUs, so that each GPU has a similar aggregated FAP value.

\subsection{Feature aggregation with one-sided reads}
\label{sec:dma}

\sys uses GPU kernels that can leverage efficient one-sided reads to access remote features over NVLink/InfiniBand. We describe how to support one-sided reads on GPUs and how to make them efficient.

\mypar{Supporting one-sided reads on GPUs} \sys supports one-sided reads on GPUs through a \emph{feature lookup table}, which converts feature IDs to their physical memory addresses on a remote device. This feature lookup table is computed when executing the feature placement algorithm, and it can be accessed efficiently by the GPU kernels through UVA. Maintaining a feature lookup table incurs a low memory overhead: the number of rows in the table grows with the number of graph nodes. Even wit a large-scale graph that has hundreds of millions of graph nodes, the table only consumes several hundreds of MBs of memory.


\mypar{Making one-sided reads efficient}
%
%
\sys uses GPU kernels with one-sided reads to access features that are sparsely distributed in memory spaces, \ie their memory locations vary because the features are randomly sampled. To increase the efficiency of one-sided reads with sparse features, \sys adopts two optimizations:

\mypari{(i)~Zero-copy optimization} \sys implements one-sided reads by leveraging the zero-copy capabilities in CPU-GPU and GPU-GPU communication. To support zero-copy access to features on a peer GPU, \sys registers the features as pinned memory using \texttt{cuda\-Host\-Register()}, which allows CUDA kernels on local GPUs to access them directly. Before reading a batch of features from registered host memory, the features are sorted according to their addresses, which leads to better locality during feature address translation on GPUs.

To support zero-copy access over InfiniBand, \sys registers the features as a memory region using \texttt{ibv\_reg\_mr()}. It then uses \texttt{ibv\_post\_send()} for one-sided RDMA reads, which avoids interrupting the CPU.  \sys allocates multiple queue pairs to parallelize RDMA reads, which improves throughput. Instead of setting the signal field and polling the completion queue for each read, \sys performs it once for each batch, which further reduces latency.

\mypari{(ii)~TLB optimization} RDMA requires address translation in the InfiniBand NIC, but random memory accesses lead to TLB misses. Assuming the features have memory addresses ranging from $2k$ to $2k+1$ on the same memory page, when reading features \eg at addresses~$<$$2, 3, 10, 11$$>$, the reading order of $<$$2, 10, 3, 11$$>$ cause 4~TLB misses, whereas the order of $<$$2, 3, 10, 11$$>$ results in only 2~TLB misses. Therefore, \sys sorts all feature reads by their memory addresses, which allows adjacent reads to be clustered together to improve TLB hit rates of the NIC.




\newcommand{\MARK}[1]{\textcolor{red}{#1}}

\section{Evaluation}
\label{sec:eval}











We evaluate the performance of \sys experimentally. \sys is written in C++, CUDA~C, and Python. It can serve GNN models written in PyG and DGL~(PyTorch). Our evaluation aims to answer the following questions:

\begin{myitemize}
\item How does \sys{}'s workload-aware approach compared to other GNN serving implementation in terms of latency and throughput?~(\S\ref{sec:end-to-end})
\item How does \sys{} scale with more GPUs and servers?~(\S\ref{sec:multiple-multigpu-servers})
\item Does \sys's PSGS metric adapt to different request ingestion rates?~(\S\ref{sec:eval:psgs})
\item Does \sys's FAP metric achieve better performance for feature access compared to existing algorithms?~(\S\ref{sec:eval_access})
\item Does \sys's one-side read strategy achieve higher throughput when collection features?~(\S\ref{sec:eval_tensors})
\item How is \sys impacted by communication links?~(\S\ref{sec:eval:comms_links})
\end{myitemize}

\subsection{Evaluation setup}


\mypar{Testbeds} We use the following hardware in our experiments: (i)~\textsf{Cluster testbed} has 3~servers, each with 2 or 4~NVIDIA A6000 GPUs (with pairwise NVLink~3.0 links) and AMD EPYC 7402P 24-core CPUs with 128\unit{GB} of host memory. The network links connections are 100-Gbps InfiniBand; and (ii)~\textsf{Cloud testbed} has 4~cloud VMs, each with 8~NVIDIA~V100 GPUs (16\unit{GB} of RAM, with NVLink in a group of 4~GPUs) and Intel Xeon Gold 5220R~(2.2\unit{GHz}) CPUs with 448\unit{GB} of host memory. The network is 10-Gbps Ethernet.


\begin{table}[t]
  \centering
  \small

\begin{tabular}{lrrc}
  \toprule
    \textbf{Dataset}  & \textbf{Nodes} & \textbf{Edges}           & \textbf{Feature size} \\
    \midrule
    ogbn-products                     &    2.45M    &      123M      &  100   \\
    ogbn-papers100M                    &     111M         &         1.6B   &     128      \\
    ogbn-mag240M                         &     240M         &          1.72B         &     768  \\
    Reddit                              &      232K        &         114M              &     300        \\
    LiveJournal                          &     4.8M         &         69M              &      N/A       \\
         ogbn-products+                     &    2.45M    &      123M      &  10000   \\
    \bottomrule
\end{tabular}

  \caption{Evaluation datasets}\label{tab:dataset} 
\end{table}

\mypar{Datasets} We use 6~public graph datasets 
(\T\ref{tab:dataset})
: (i)~\textsf{ogbn-products}~\cite{ogb}, a medium graph with product relations at Amazon; (ii)~\textsf{Reddit}, a medium graph of social communities; (iii)~\textsf{ogbn-papers100M}, a large graph of paper citation networks; (iv)~\textsf{ogbn-mag240M}, a large graph of paper citation networks; (v)~\textsf{Live Journal}, a medium graph of journal communities; and (vi)~\textsf{ogbn-products+}, the Amazon product graph extended to have large features, matching our production workloads.


\mypar{GNN models} We choose 2~popular GNN models: (i)~\textsf{GraphSAGE}~\cite{graphsage}, with k-hop neighbour sampling without replacement (hidden dimension of 256); and (ii)~Graph Attention Network~(\textsf{GAT})~\cite{gat} with 4 attention heads. For both models, we use a local batch size of $1024$ on each worker. All models are implemented using PyG and PyTorch. We also evaluated other GNN models (GraphSaint~\cite{graphsaint} and ClusterGCN~\cite{clustergcn}), observing similar results (omitted due to space limitations).

\mypar{Baselines} We compare \sys against two state-of-the-art GNN systems, \textsf{PyG}~\cite{pyg} (v2.0.1) and \textsf{DGL}~\cite{dgl} (v0.7.0). We extend these systems to process GNN requests (\ie using the test mode of PyG and DGL). Since \sys supports models imported from PyG and DGL, we adopt their task implementations (\eg for graph sampling, DNN inference) when possible. This way, performance differences can be attributed to the different request processing and feature placement.

Note that \textsf{PaGraph}~\cite{pagraph}, \textsf{BGL}~\cite{bgl} and \textsf{GNNLab}~\cite{gnnlab} support training only, and we could not extend them to supporting GNN serving. However, we re-implement the proposal by PaGraph for feature placement. Since \textsf{P3}~\cite{p3} is not open-source, we also re-implement its approach in \sys, reproducing its published performance results. While we exclude AliGraph~\cite{aligraph} from our end-to-end experiments, because it does not support PyTorch, we implement its feature placement approach.


\mypar{Request workload} We launch  multiple client processes that continuously produce GNN requests. Each request randomly samples input nodes with the out-degree as the weight, which is representative of real-world serving workloads. 

\subsection{Throughput and latency}
\label{sec:end-to-end}


To assess end-to-end performance, we measure the throughput and latency of serving a GNN model for a given dataset. We evaluate three scenarios: (i)~users want the highest possible throughput with a given latency target; (ii)~users want the lowest possible latency; and (iii)~users want high throughput when serving large GNN models.



\begin{figure}[t]
\centering
\includegraphics[width=0.87\linewidth]{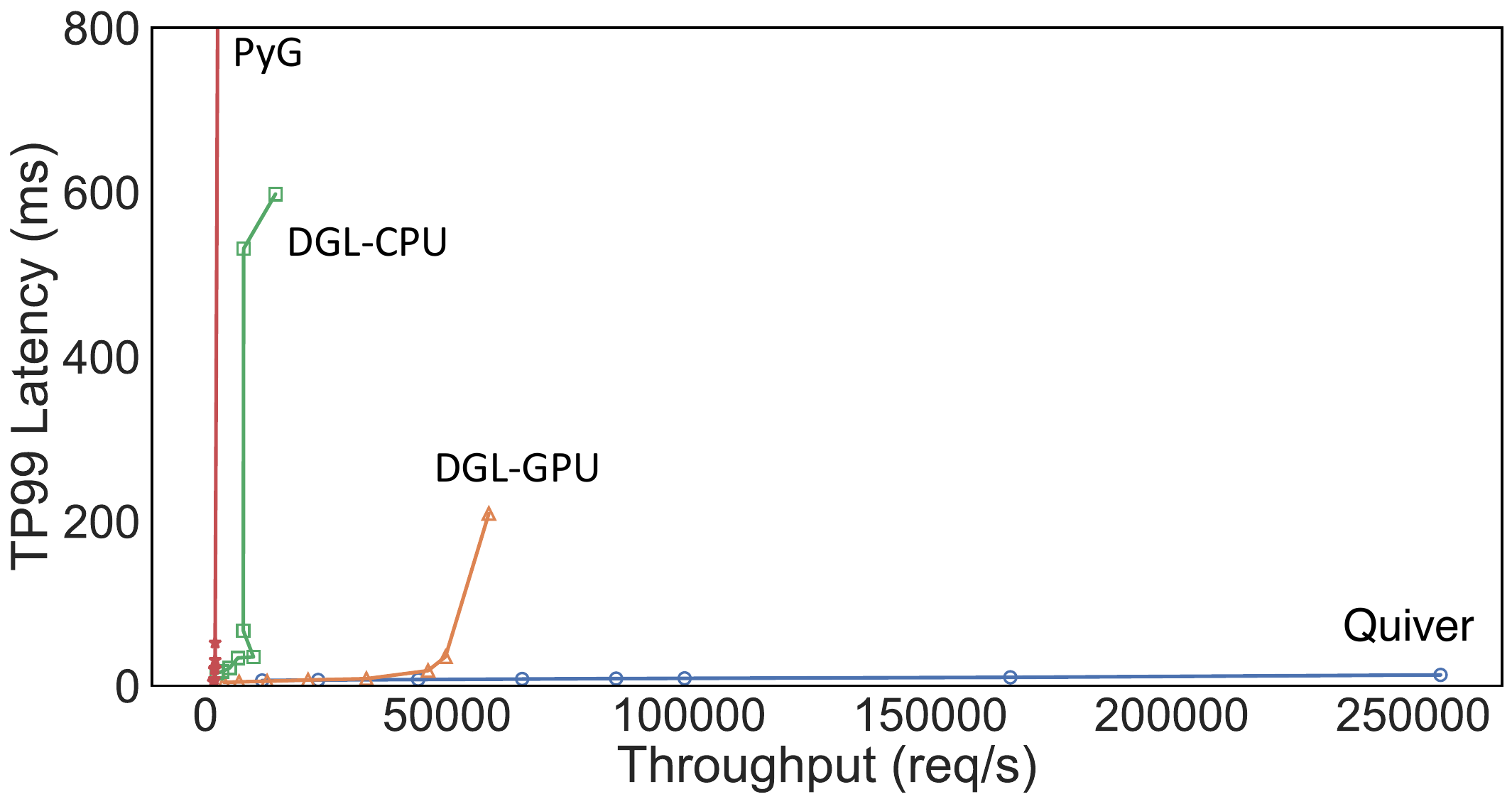}
\caption{Throughput vs. latency of GNN request serving
}
\label{fig:tp99}
\end{figure}

\mypar{Throughput vs.\ latency}
\label{sec:throughput-vs-latency}
First, we compare \sys with PyG and DGL (with both CPU and GPU sampling) in terms of throughput and latency running on one server with 2 GPUs from the \textsf{cluster testbed}. We vary the batch size from 8 to 1024 to generate different scale workload and record the throughput and 99\textsuperscript{th} latency percentile.
\F\ref{fig:tp99} shows the throughput/latency plot when processing GNN requests. We observe that the PyG's latency increases substantially with a higher throughput to over 1\unit{sec}. DGL with CPU sampling behaves similarly, but DGL with GPU sampling achieves a higher throughput of just above 50,000\unit{reqs/sec}. In contrast, \sys maintains latencies below 13\unit{ms}, despite processing requests at a peak throughput of 255,000\unit{reqs/sec} , when we have reached system full load, with CPU utilization at 95-100\% and GPU utilization at 80-85\%. Since \sys only allocates sampling tasks that benefit from GPU processing to GPUs, while avoiding data movement bottlenecks between GPUs, it achieves a substantially higher throughput without a latency penalty. 


\begin{figure}[t]
\centering
\includegraphics[width=0.4\textwidth]{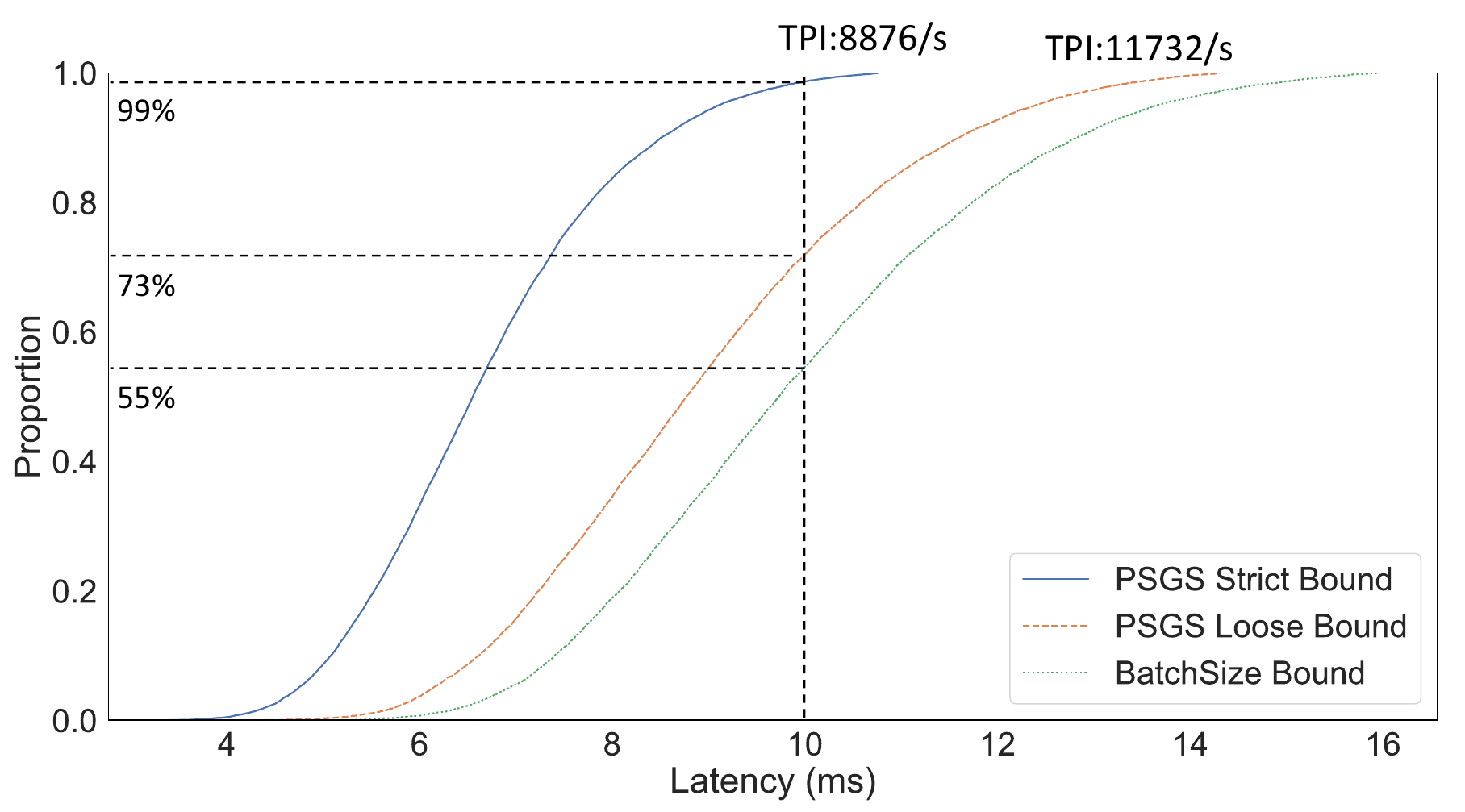}
\caption{Latency with different policies}
\label{fig:latency-bound}
\end{figure}

\mypar{Strict vs. loose latency bounds} \sys supports two settings for latency targets using the PSGS metric: \textsf{PSGS-Strict}, which apply upbound line to present the relationship between latency and PSGS, it strictly achieves a given latency bound; while \textsf{PSGS-Loose}, which use average line, it focuses on high throughput with a relaxed latency bound. We compare \textsf{PSGS-Strict}, \textsf{PSGS-Loose}, and a fixed batch size (\textsf{Batchsize-Bound}) as a baseline. We set the target 99\textsuperscript{th} percentile latency to 10\unit{ms} for PSGS and a fixed batch size that the most serves requests below 10\unit{ms}.
\F\ref{fig:latency-bound} shows the CDF plot of the latency of \textsf{PSGS-Strict}, \textsf{PSGS-Loose} and \textsf{Batchsize-Bound}, which handle 99\%, 73\%, and 55\% of queries within 10\unit{ms}, respectively. While having a higher latency bound, \textsf{PSGS-Loose} maintains higher throughput (11,700\unit{reqs/sec}), which is 57\% higher than PSGS-Strict's throughput (8,800\unit{reqs/sec}). This shows the flexibility of using the workload-aware PSGS metric, allowing it to be adjusted to different scenarios.

\label{sec:single-multigpu-server}


\subsection{Scalability}
\label{sec:multiple-multigpu-servers}

\begin{figure}[t]
\centering
  \begin{subfigure}{.23\textwidth}
    \includegraphics[width=\linewidth]{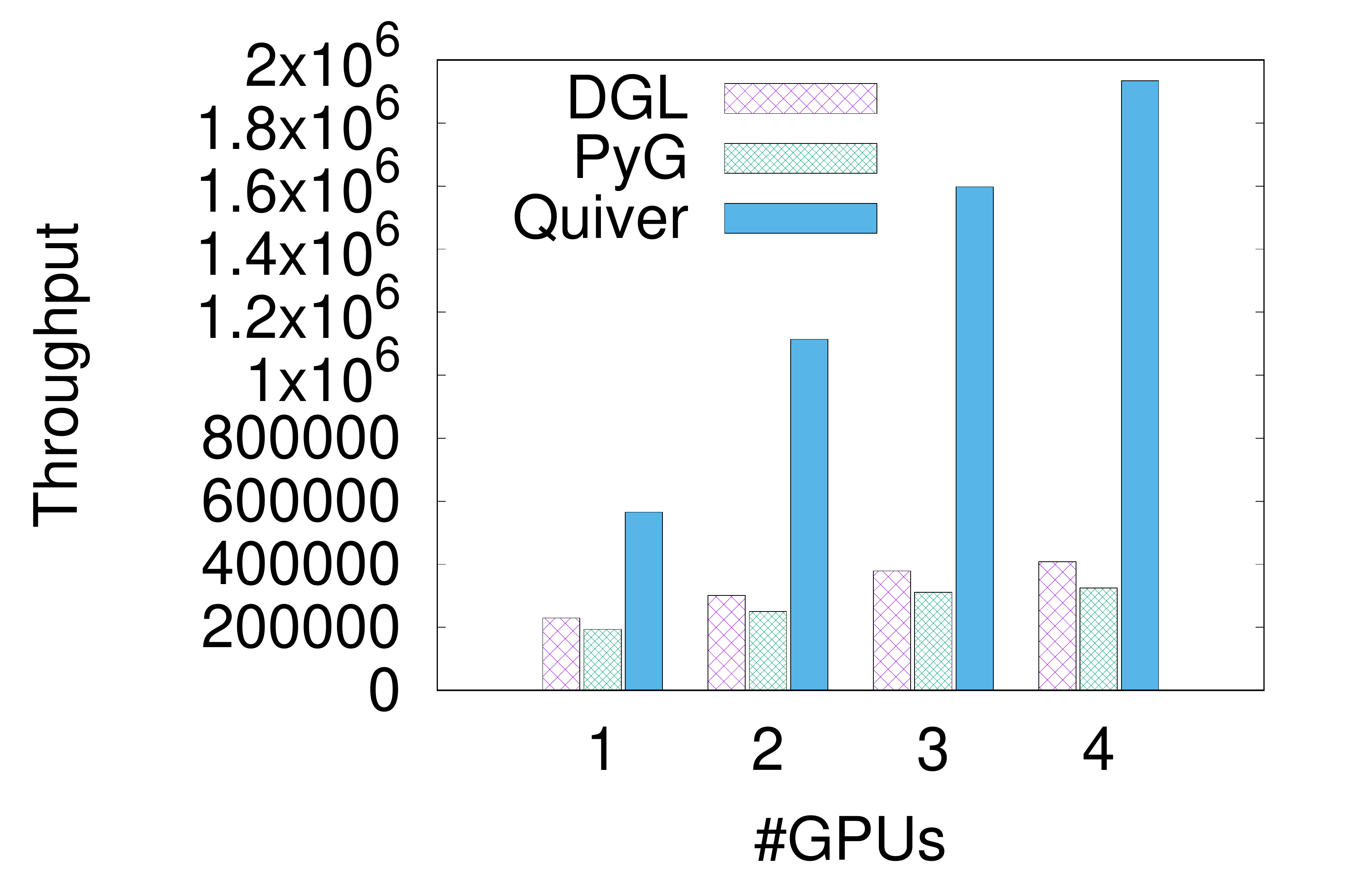}
    \caption{ogbn-products}
    \label{fig:single-server-e2e-ogbn}
  \end{subfigure}
  \begin{subfigure}{.23\textwidth}
    \includegraphics[width=\linewidth]{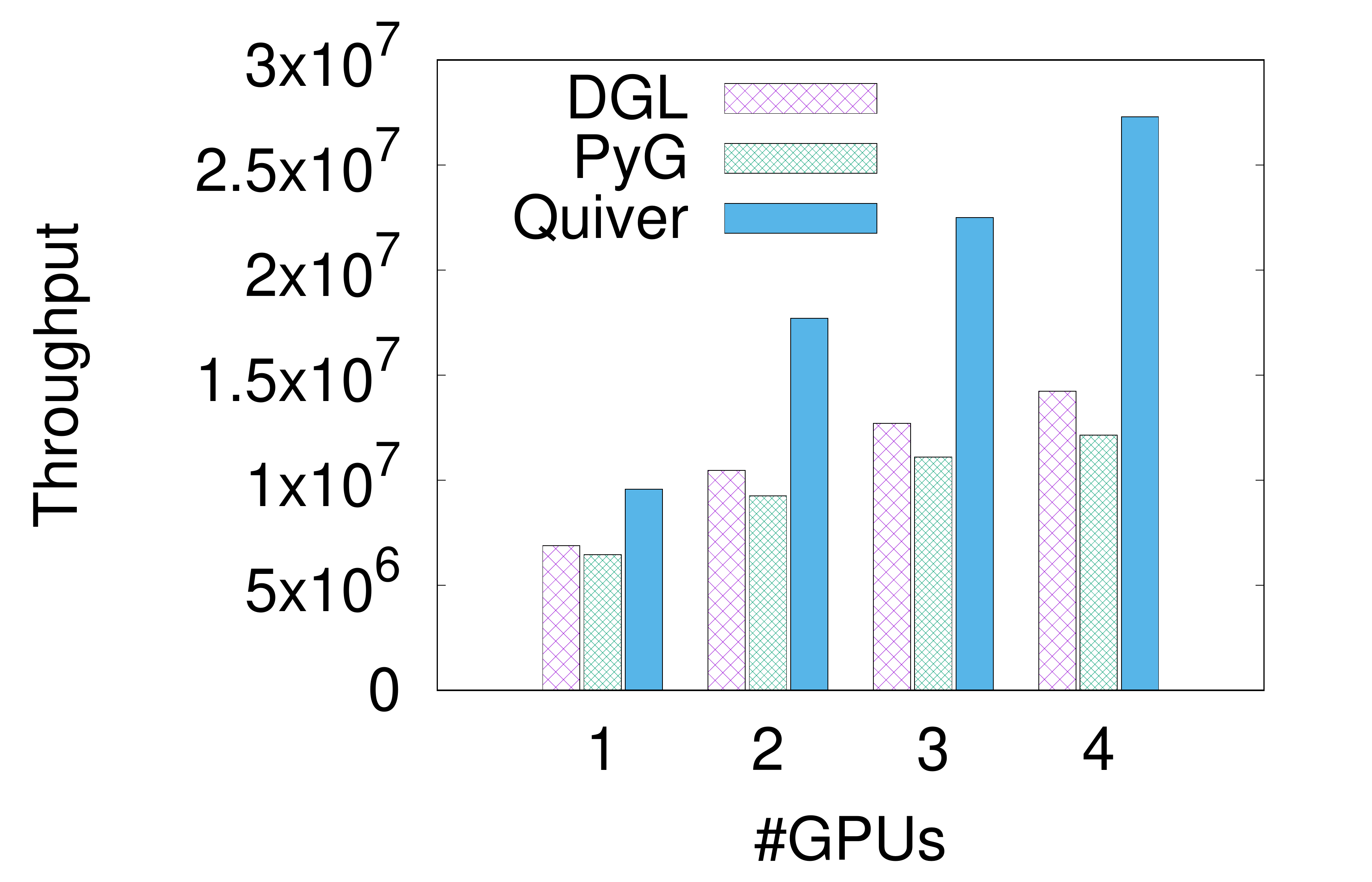}
    \caption{ogbn-papers100M}
    \label{fig:single-server-e2e-papers}
  \end{subfigure}
  \caption{Scalability with single multi-GPU server}
  \label{fig:single-server-e2e}
\end{figure}

Next we evaluate \sys{}'s scalability. We increase the throughput until the systems reach a user latency threshold of 30\unit{ms}. We then report the achieved maximum throughput. 

\mypar{Single server} We first explore how well \sys scales to multiple GPUs in a single server in our \textsf{cluster testbed} when serving the \textsf{GraphSage} model. \F\ref{fig:single-server-e2e} shows the achieve throughput with an increasing number of GPUs compared to PyG and DGL: with a small dataset (\textsf{ogbn-products} in \F\ref{fig:single-server-e2e-ogbn}), \sys handles 570,000\unit{reqs/sec} using a single GPU -- in contrast, \textsf{DGL} and \textsf{PyG} achieves 220,000\unit{reqs/sec} (3.8$\times$ fewer) and 200,000\unit{reqs/sec} (4.7$\times$ fewer), respectively. \sys benefits from exploiting multiple pipelines per GPU and its efficient one-sided reads; with 2--4 GPUs, \sys is 8.3$\times$ and 10.1$\times$ faster than DGL and PyG, respectively. By caching features across GPUs based on the FAP metrics, \sys achieves substantially higher throughput. 



For the \textsf{paper100M} dataset (\F\ref{fig:single-server-e2e-papers}), when running with 4~GPUs, \sys benefits from larger total amount of GPU memory, which enables it to schedule more work to the GPUs. As a consequence, \sys is 3.2$\times$ and 3.9$\times$ faster than DGL and PyG, respectively. 


\begin{figure*}[tb]
  \centering
    \begin{subfigure}{.24\textwidth}
      \includegraphics[width=\linewidth]{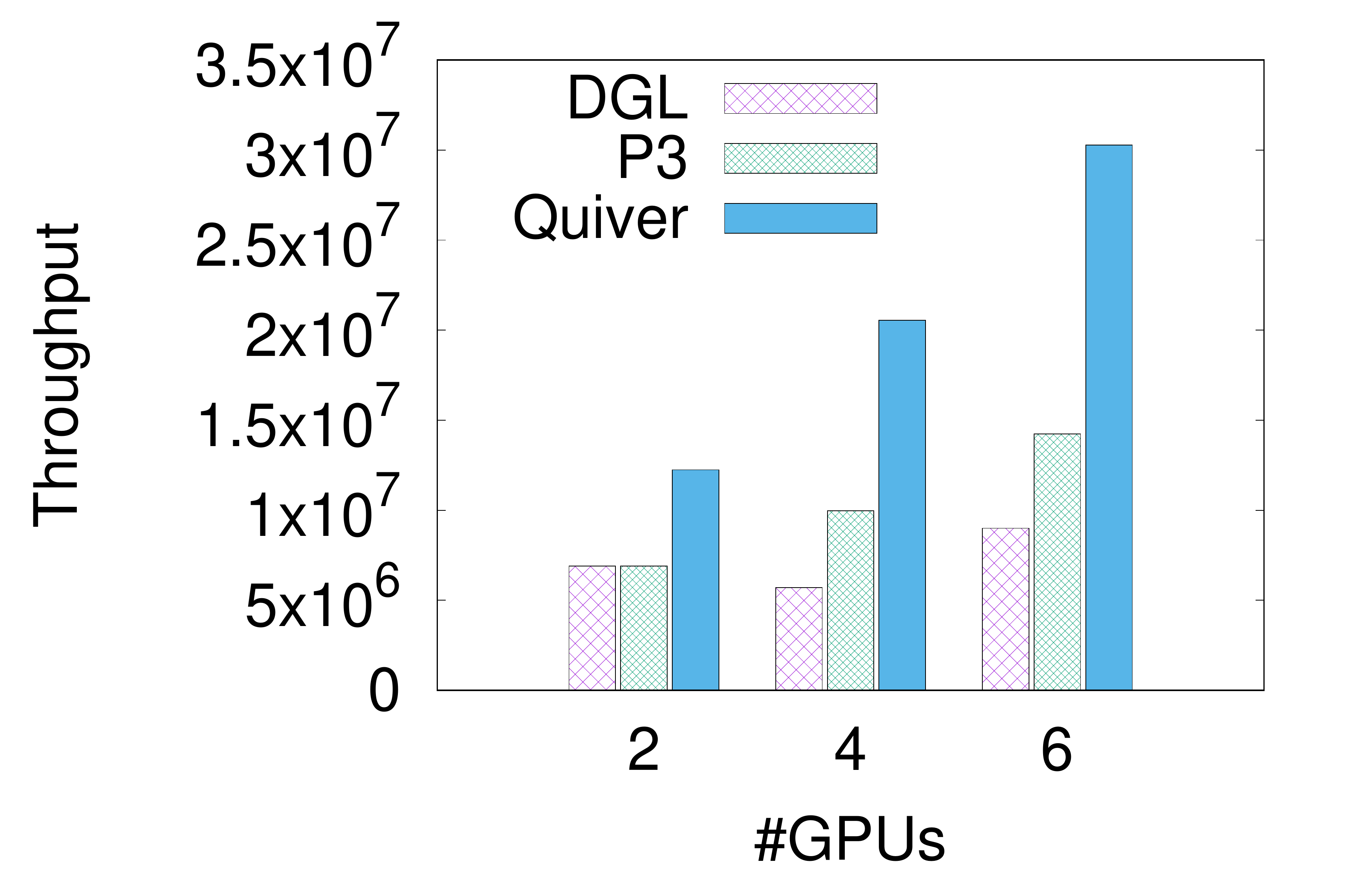}
      \caption{\label{fig:multi-server-paper100m-gat}Paper100M/GAT}
    \end{subfigure}
    \begin{subfigure}{.24\textwidth}
      \includegraphics[width=\linewidth]{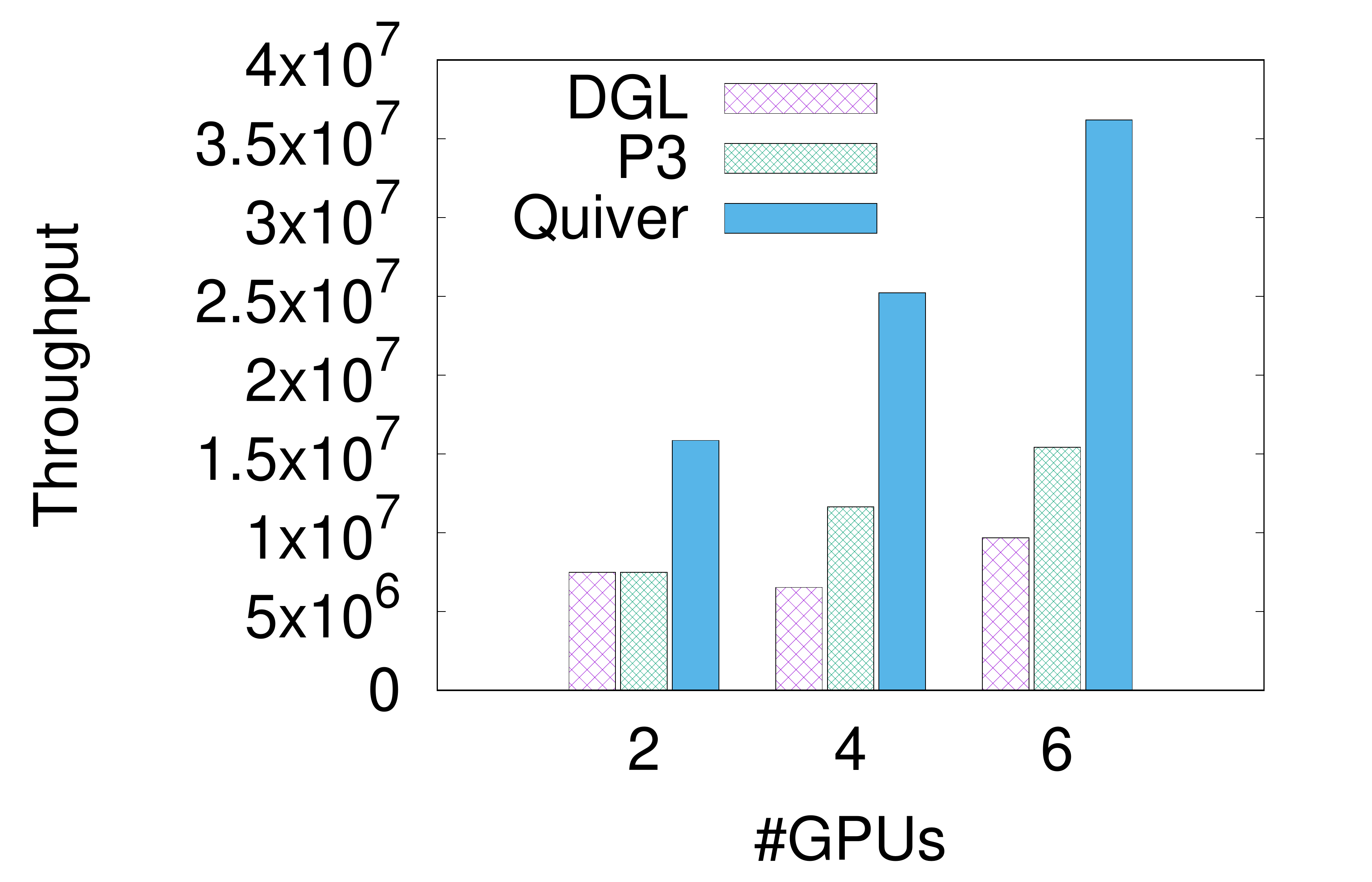}
      \caption{\label{fig:multi-server-paper100m-sage}Paper100M/GraphSage}
    \end{subfigure}
    \begin{subfigure}{.24\textwidth}
      \includegraphics[width=\linewidth]{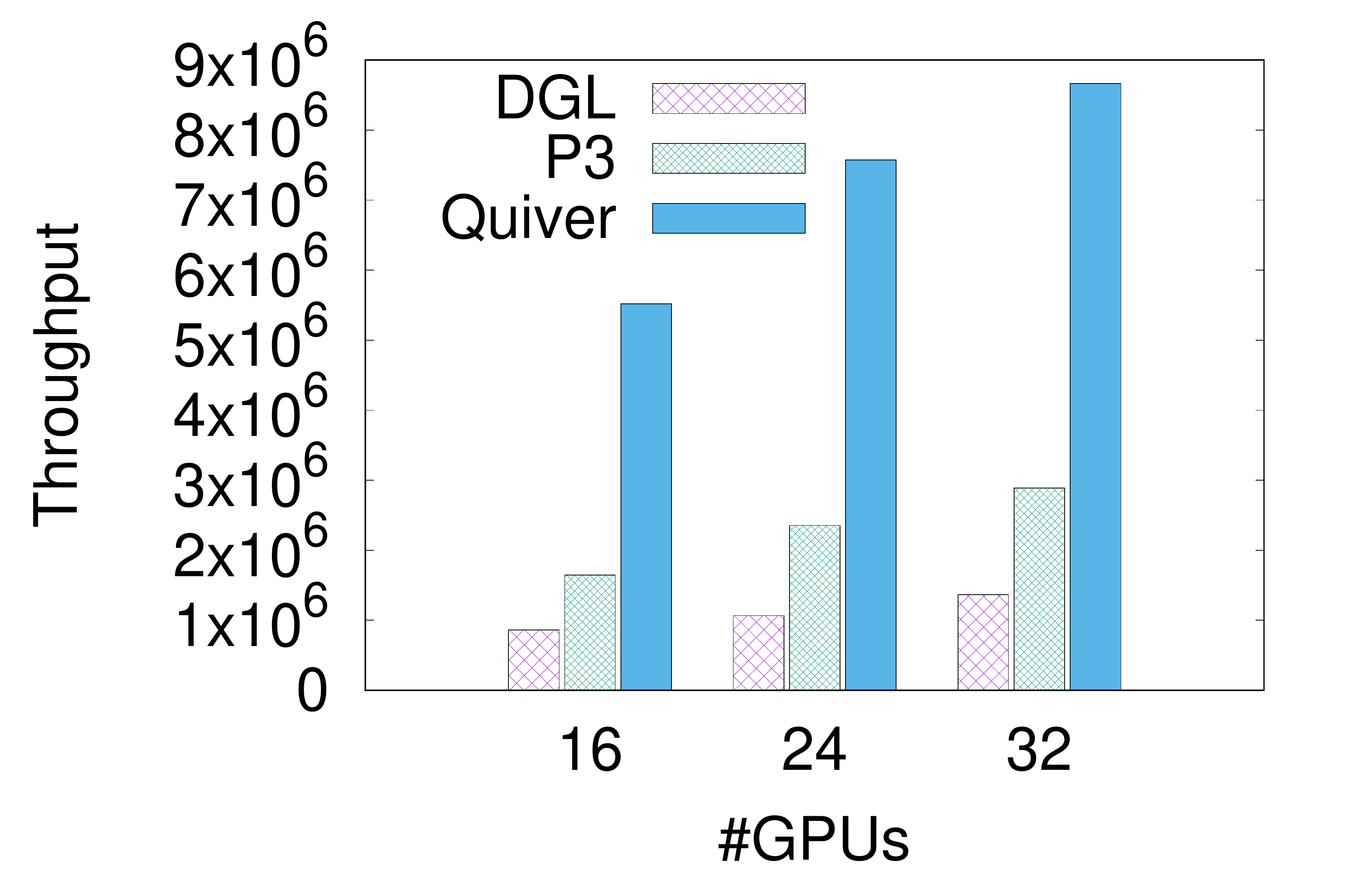}
      \caption{\label{fig:multi-server-mag240m-gat}Mag240M/GAT}
    \end{subfigure}
    \begin{subfigure}{.24\textwidth}
      \includegraphics[width=\linewidth]{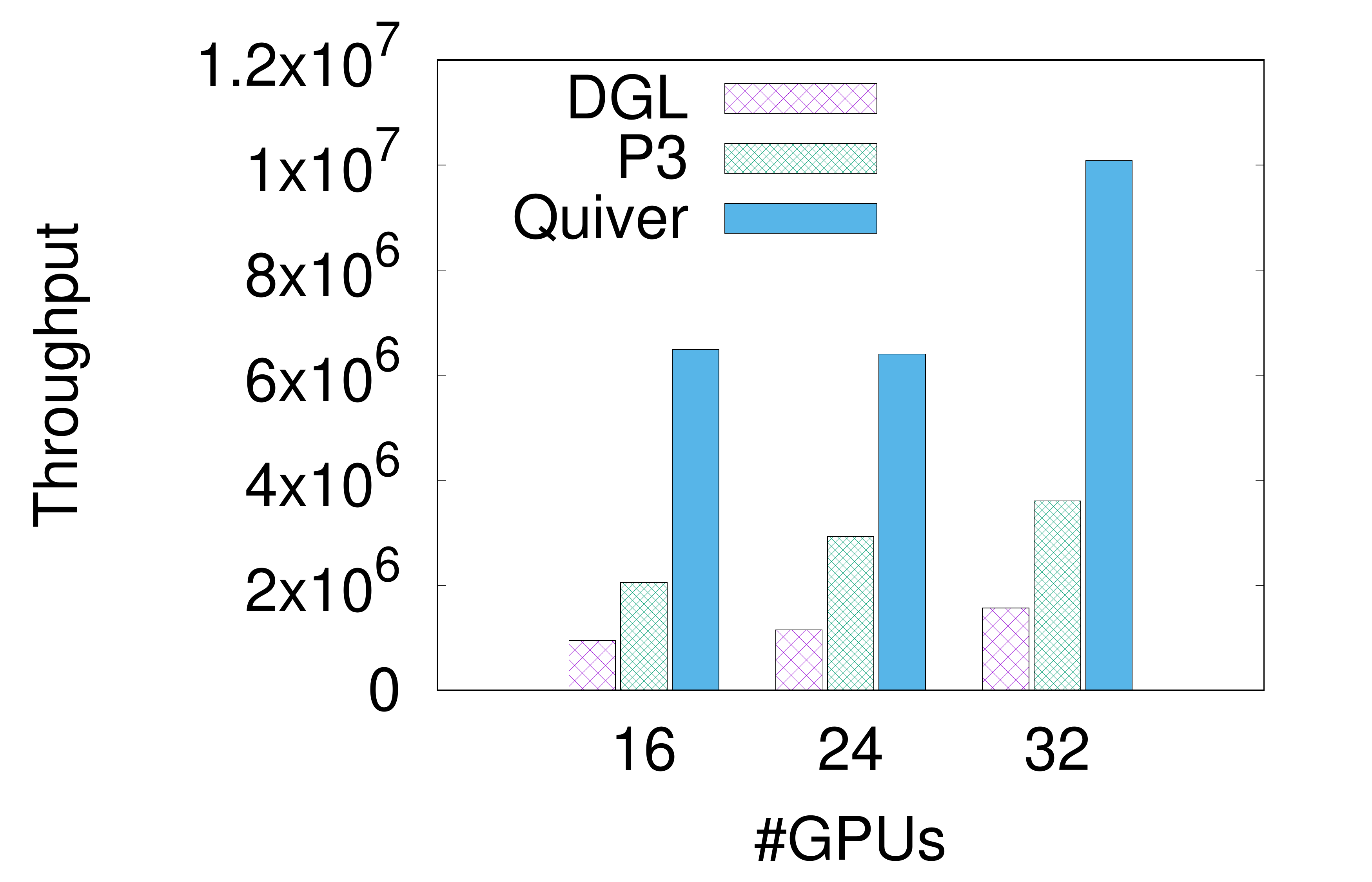}
      \caption{\label{fig:multi-server-mag240m-sage}Mag240M/GraphSage}
    \end{subfigure}
    \caption{Scalability with multiple multi-GPU servers}
    \label{fig:multi-server}
  \end{figure*}




\mypar{\textsf{Cluster testbed}} We use the \textsf{cluster testbed} with 3~servers (2~NVIDIA~A6000 GPUs each). The 3~servers have sufficient memory to fit the \textsf{paper100M} dataset. DGL and P3 use their default strategies to partition the \textsf{paper100M} dataset.
For \textsf{GAT} on a single server (2~GPUs) (see~\F\ref{fig:multi-server-paper100m-gat}), \sys achieves a 2.8$\times$ speedup over both DGL and P3. With 2~servers (4~GPUs), P3 has better scalability compared to DGL, because it reduces the features sent over the network. P3 is thus 1.8$\times$ faster than DGL, which is consistent with the results from the P3 paper~\cite{p3}. In contrast, \sys achieves better scalability (3.2$\times$) than P3. Since \sys replicates graph data and features in GPU memory, it can reduce communication costs. It also accounts for InfiniBand connectivity when deciding between partitioning and replication. \sys{}'s throughput improves with more servers: with 3~servers (6~GPUs), it is 3.4$\times$ and 7.1$\times$ faster than P3 and DGL, respectively.
For \textsf{GraphSage} (see~\F\ref{fig:multi-server-paper100m-sage}), \sys achieves better results compared to GAT: with 3~servers (6~GPUs), it manages a 4.8$\times$ and 8.4$\times$ speedup compared to P3 and DGL, respectively. GraphSage places a strong emphasis on graph sampling and uses a smaller GNN model compared to GAT. This means that \sys can more effectively optimize its feature placement, \eg by placing more graph data and features in GPU memory due to the smaller GraphSage GNN size. It also increases the benefits of multiplexing GPU pipelines, \eg by executing more feature aggregation, sampling tasks and DNN inference tasks on GPUs.

\mypar{\textsf{Cloud testbed}} On the more powerful \textsf{cloud testbed} with up to 32 GPUs, we use the \textsf{mag240M} dataset, which is the largest GNN dataset in the OGB benchmark. With 2~servers (16~GPUs) (see~\F\ref{fig:multi-server-mag240m-gat}), \sys achieves 5.5$\times$ and 2.8$\times$ the througput of DGL and P3, respectively, when serving the GAT model. With 4~servers (32~GPUs), \sys improves the speedup ratios to 7$\times$ and 3.2$\times$, respectively. The same behavior can be also seen with \textsf{GraphSage}: with 32 GPU, \sys achieves speedups of 7.9$\times$ and 3.3$\times$ compared to DGL and PyG, respectively. Note that these speed-up ratios are larger than those in the 16-GPU case.

\sys{}'s performance improvement grows with the number of GPUs (or servers), because it fully utilizes the available CPU/GPU memory. With more servers, the CPU and GPU memory increases, but it leads to more intra-server communication. As a result, \sys replicates more frequently-accessed feature data to improve locality, reducing the impact of these communication overheads. In contrast, DGL and P3 cannot fully exploit all cluster memory, and their scalability becomes limited by these network bottlenecks.


\label{sec:micro}


\subsection{Robustness to data skew}
\label{sec:eval:psgs}

\begin{figure*}[tb]
    \centering
      \begin{subfigure}{.23\textwidth}
        \includegraphics[width=\linewidth]{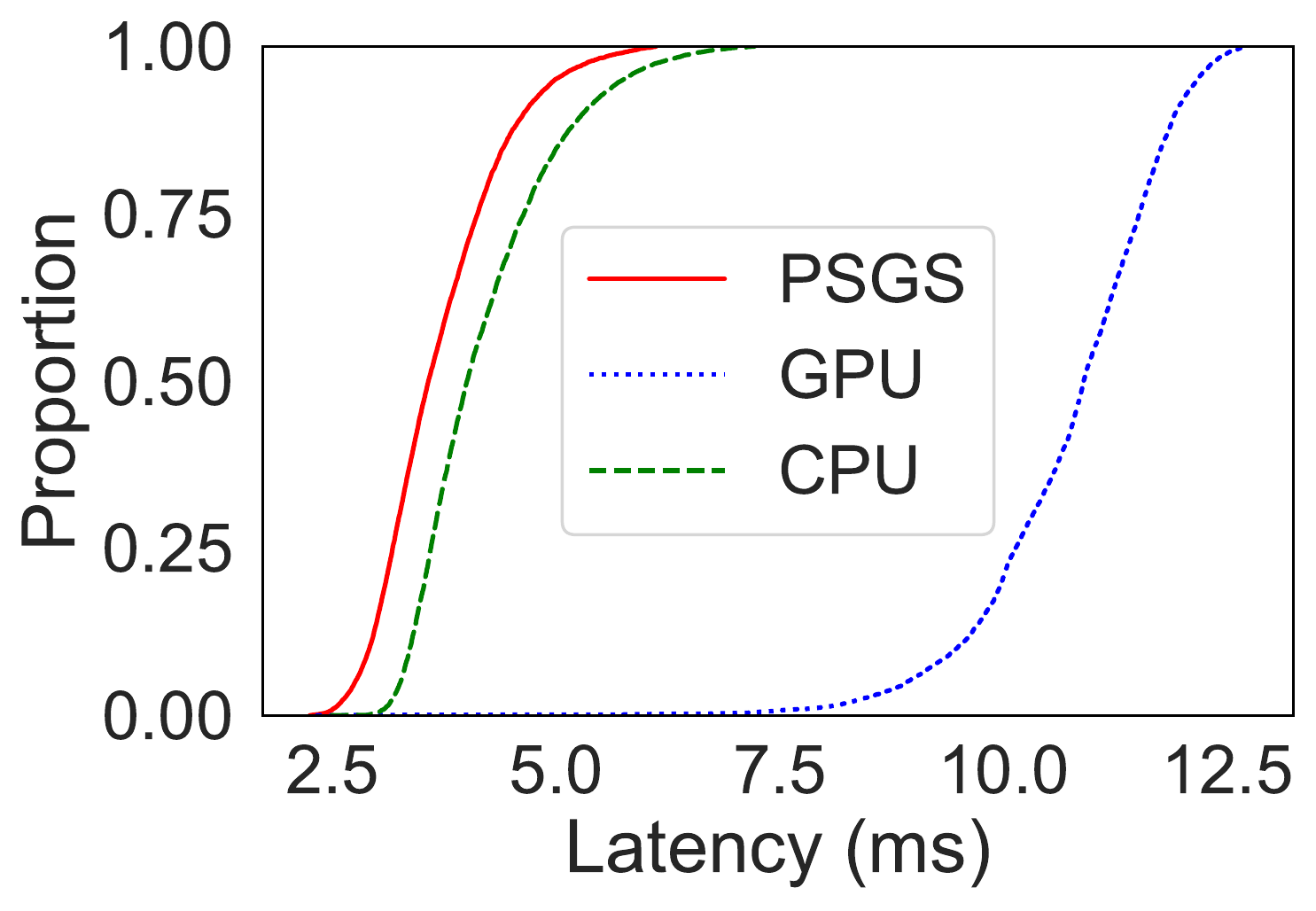}
        \caption{\label{fig:irr-small}Small workload}
      \end{subfigure}
      \begin{subfigure}{.23\textwidth}
        \includegraphics[width=\linewidth]{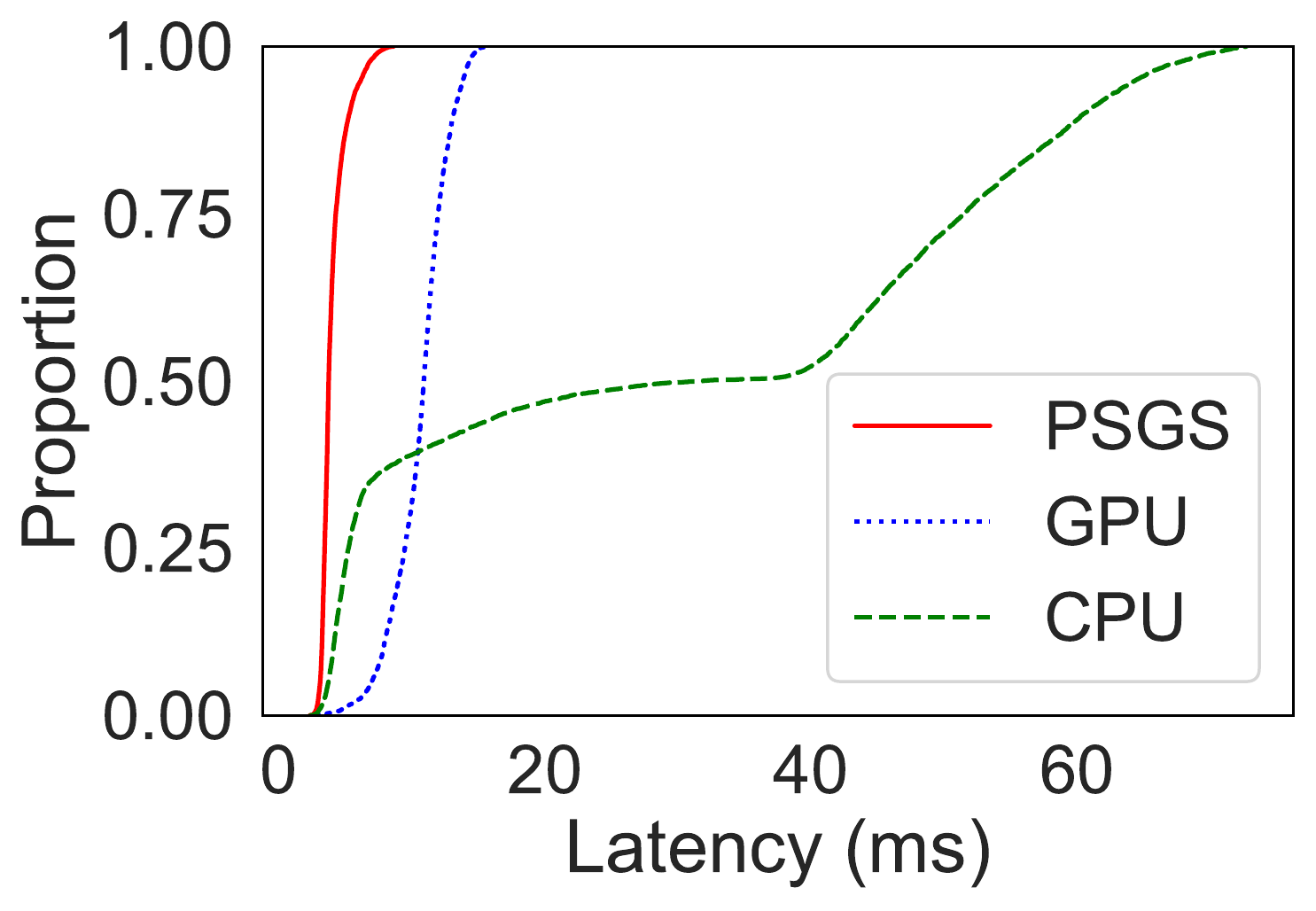}
        \caption{\label{fig:irr-medium}Medium workload}
      \end{subfigure}
      \begin{subfigure}{.23\textwidth}
        \includegraphics[width=\linewidth]{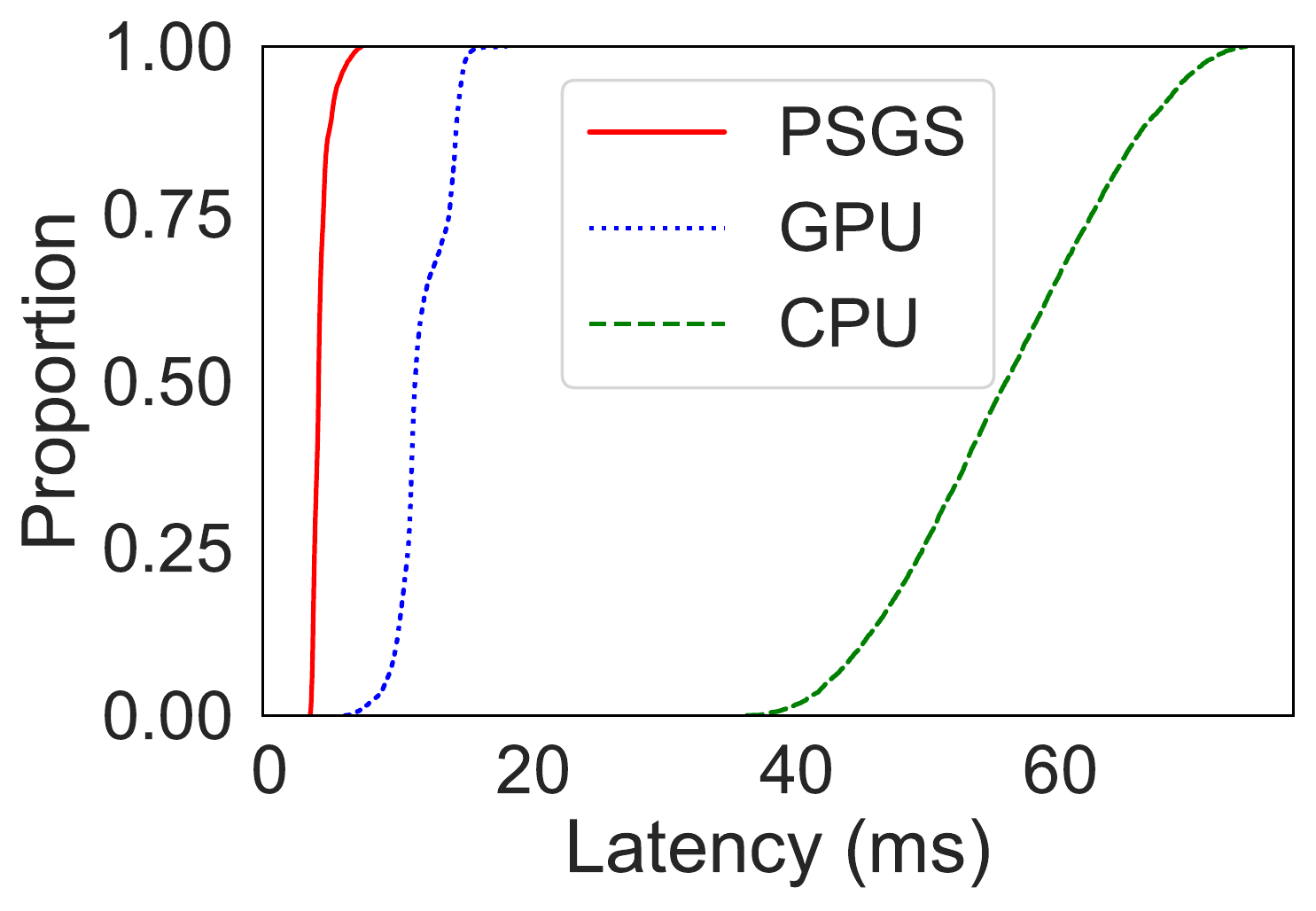}
        \caption{\label{fig:irr-large}Large workload}
      \end{subfigure}
      %
      \caption{Different sampling strategies with skewed datasets}
      \label{fig:irr}
    \end{figure*}
  
\begin{figure}[t]
  \centering
  \begin{subfigure}{.23\textwidth}
    \includegraphics[width=\linewidth]{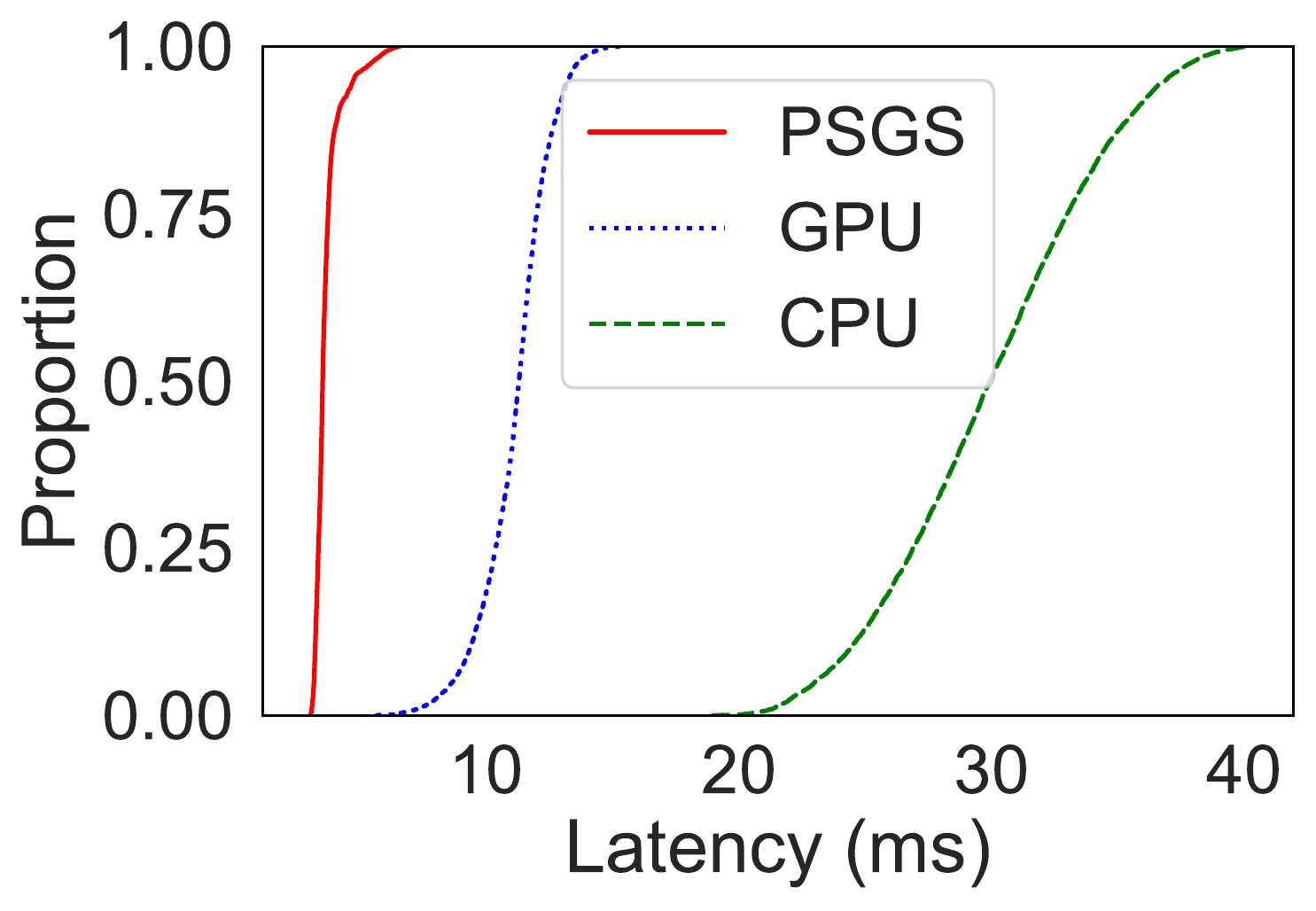}
    \caption{\label{fig:max-batch} Large batch size}
  \end{subfigure}
  \begin{subfigure}{.23\textwidth}
    \includegraphics[width=\linewidth]{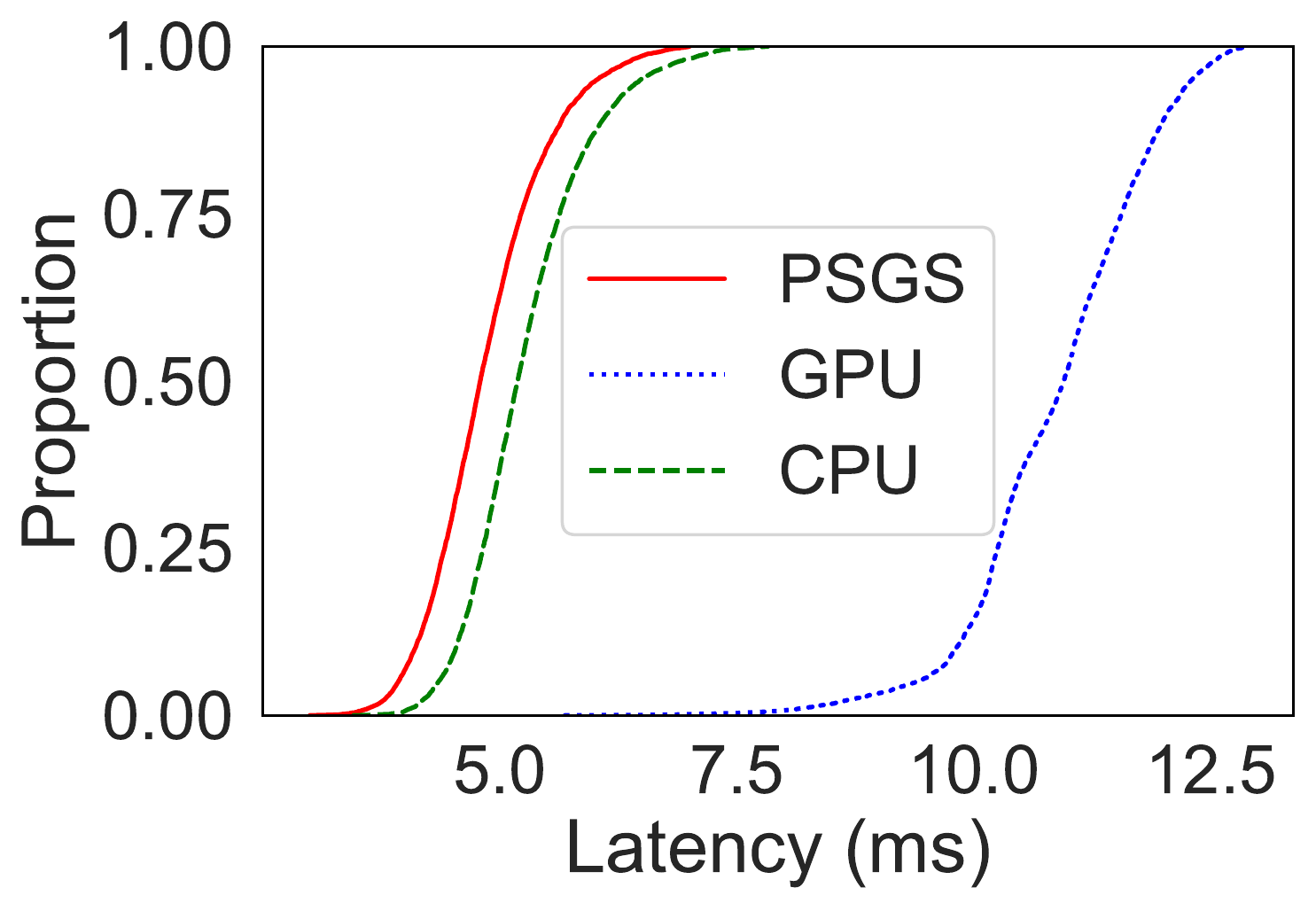}
    \caption{\label{fig:min-batch} Small batch size}
  \end{subfigure}
  \caption{Comparison of random sampling strategies
  }
  \label{fig:normal}
\end{figure}

In this experiment, we investigate if \sys's PSGS metric yields the best performance when facing irregular input data. We use the \textsf{reddit} dataset and a 2-layer \textsf{GraphSAGE} model. The fan-out of each layer is set to 25 and 10, respectively. We use the \textsf{cluster testbed} with 2~NVIDIA~A6000 GPUs. We compare three strategies: (i)~our workload-aware \textsf{PSGS} strategy for sampling; (ii)~\textsf{CPU}-based sampling; and (iii)~\textsf{GPU}-based sampling. For each strategy, we use a batch size of 96 to construct the initial nodes and then perform neighbor sampling on the dataset to obtain different workloads: for example, we select nodes with high degrees and low degrees as seeds in the \textsf{large} and \textsf{small} workloads, respectively. The \textsf{small} workload contains 4$\times$; the \textsf{medium} workload contains 170$\times$; and the \textsf{large} workload contains 280$\times$ the initial nodes.
\F\ref{fig:irr} shows that the \textsf{PSGS} strategy achieves the best performance in all cases: with the \textsf{large} workload, the GPU-based strategy performs better than the CPU-based strategy. It utilizes the GPU's ability for high throughput/low latency computation for sampling; with the \textsf{small} workload, the CPU-based strategy performs better, because it can reduces the overhead of data transfers between CPUs and GPUs.
We also compare the performance of the strategies with different batch sizes. We use a \textsf{small} batch size of 4 and a \textsf{large} batch size of 96. We randomly sample batches and perform two-layer neighbour sampling. As \F\ref{fig:irr} shows, we observe the same trend as in \F\ref{fig:irr}: \textsf{PSGS} achieves the best performance irrespective of the batch size: with \textsf{large} batches, the GPU-based strategy performs better than the CPU-based strategy; with \textsf{small} batches, the CPU-based strategy performs better than the GPU-based strategy.





\subsection{Effectiveness of feature placement}
\label{sec:eval_access}


\begin{figure}[t]
    \centering
    \begin{subfigure}{.22\textwidth}
        \includegraphics[width=\linewidth]{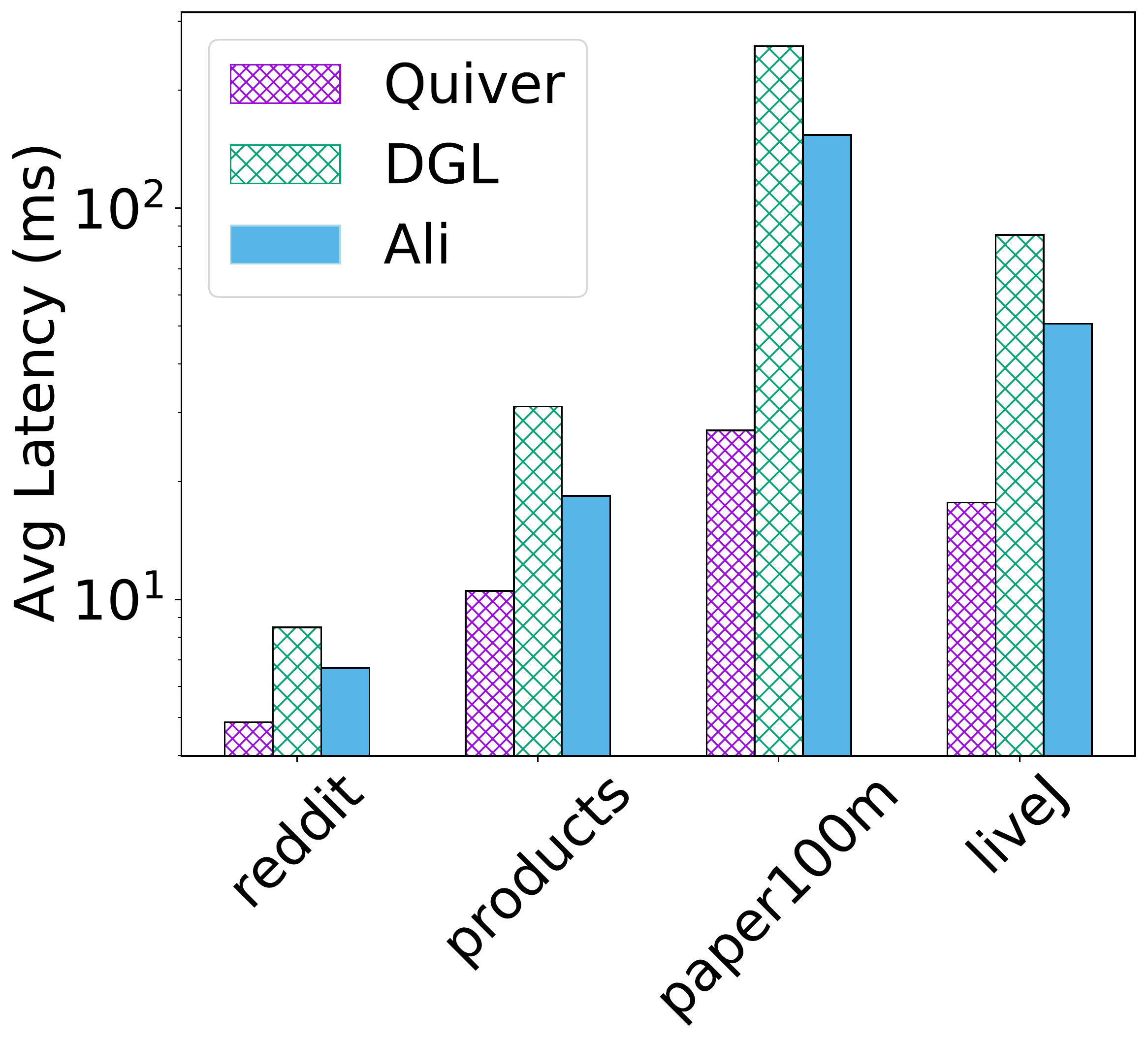}
        \caption{2~servers}\label{fig:2machine} 
    \end{subfigure}
    \begin{subfigure}{.22\textwidth}
        \includegraphics[width=\linewidth]{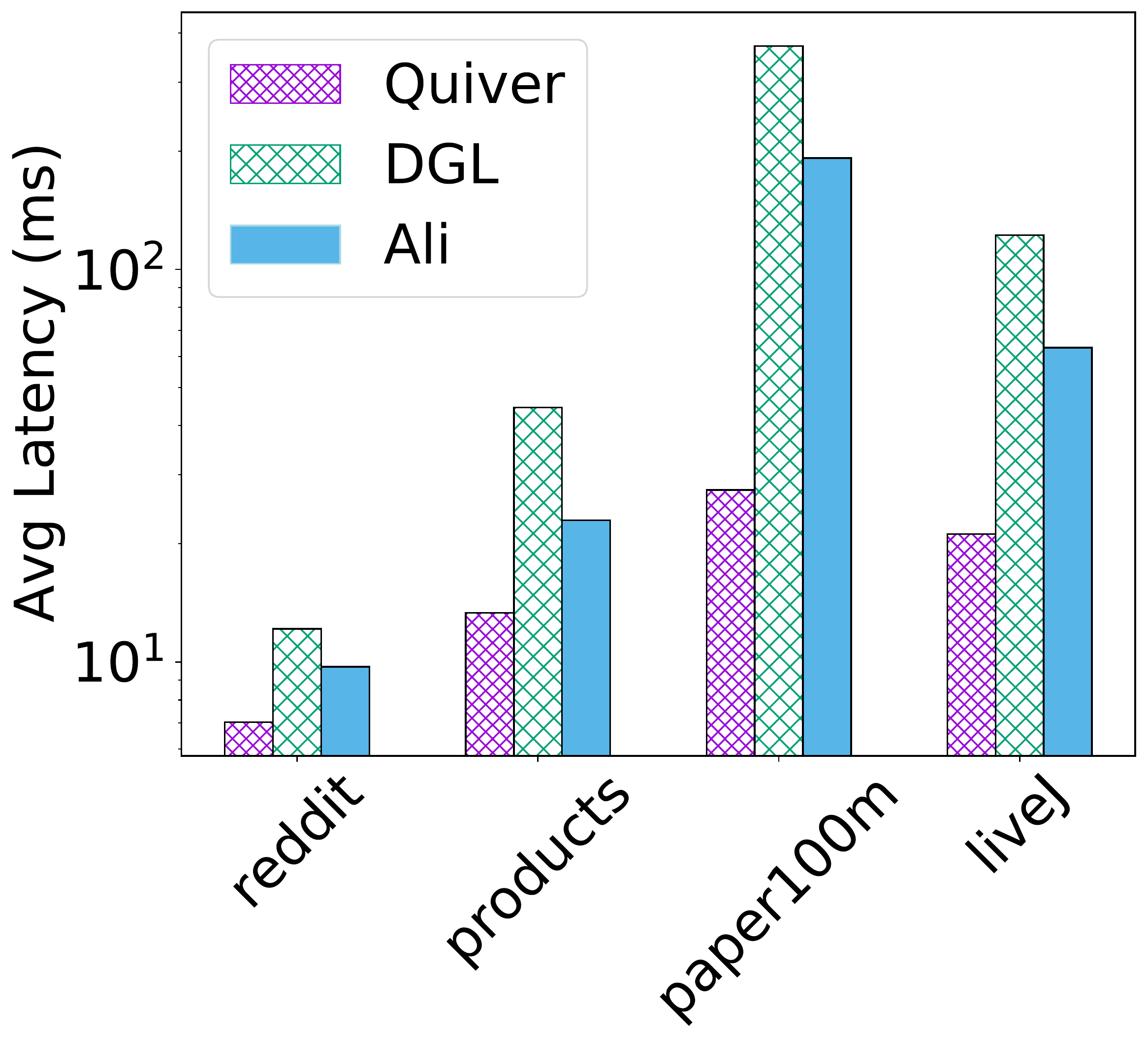}
        \caption{8~servers}\label{fig:8machine} 
    \end{subfigure}
    \caption{Latency impact of feature placement}
    \label{fig:latency-compare}
\end{figure}

We evaluate \sys{}'s workload-aware feature placement using the FAP metric on 2 and 8 servers. We compare against two baselines using 4~datasets (\textsf{Reddit}, \textsf{ogbn-products}, \textsf{ogbn-papers100m}, \textsf{LiveJ}): (i)~hash-based graph partitioning, which is the default for DGL; and (ii)~importance-based graph partitioning, which is used by AliGraph~\cite{aligraph}. The latter considers the degrees of graph nodes and performs a balanced graph cut, which is similar to Metis~\cite{metis}. We allow all devices to have 20\% extra memory to replicate data. DGL uses halo nodes to cache hot data; AliGraph uses an LRU cache for recently accessed data.
\F\ref{fig:latency-compare} shows that \sys outperforms DGL and AliGraph in terms of serving latency across all datasets and platforms: on the \textsf{Reddit} dataset, \sys has a serving latency of 4.9\unit{ms} and 7.0\unit{ms} on 2 and 8~servers, respectively; DGL and AliGraph achieve latencies of 8.5\unit{ms} and 6.7\unit{ms} on 2~machines, and 12.2\unit{ms} and 9.7\unit{ms} on 8 machines, respectively.
As the number of servers increases from 2 to 8, the serving latency increases for all platforms and datasets. This can be attributed to the increased communication overhead in a distributed setting: with 8~devices, as shown in \F\ref{fig:8machine}, the performance of hash-based partitioning (DGL) quickly degrades (\eg for \textsf{paper100m}, the latency grows from 259\unit{ms} to 370\unit{ms}) because it is workload-agnostic. The performance of AliGraph also slightly decreases (\eg for \textsf{paper100m}, the latency grows from 153\unit{ms} to 192\unit{ms} compared to the 2-device case). In contrast, \sys sustains a low latency that is much lower than AliGraph across all datasets. We speculate that \sys{}'s performance improvement over DGL and AliGraph will become even more significant for larger deployments due to its more accurate estimation of data access probabilities and its use of replication.

\vspace{-2mm}
\subsection{Performance of feature collection}
\label{sec:eval_tensors}


\begin{figure}
\centering
  \begin{subfigure}{.23\textwidth}
    \includegraphics[width=\linewidth]{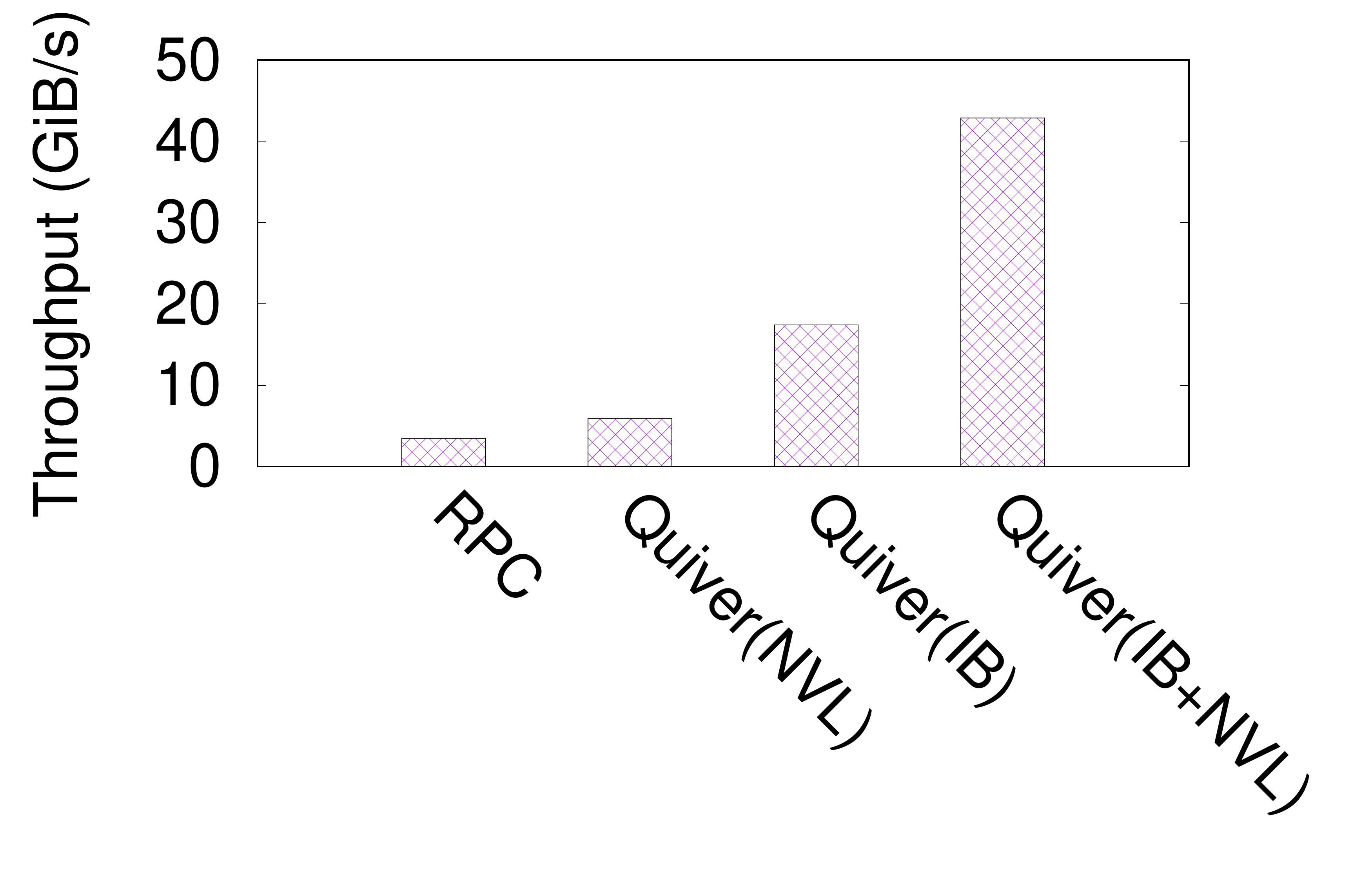}
    \caption{\label{fig:unified-tensor-paper}Paper100M}
  \end{subfigure}
  \begin{subfigure}{.23\textwidth}
    \includegraphics[width=\linewidth]{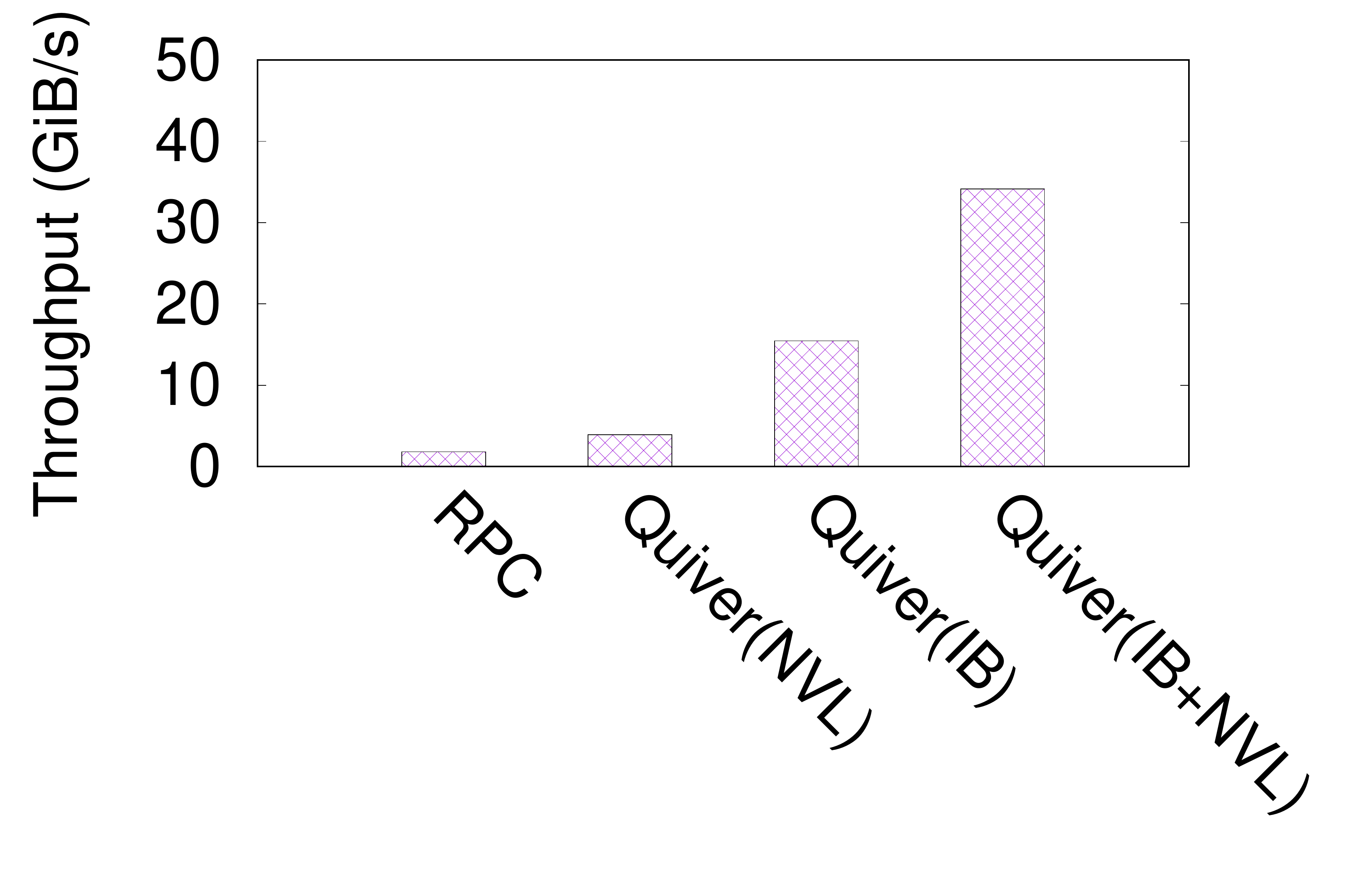}
    \caption{\label{fig:unified-tensor-mag}Mag240M}
  \end{subfigure}
  \caption{Throughput of feature collection}\label{fig:unified-tensor}
\end{figure}

Finally, we evaluate the performance of feature collection in \sys. We measure the throughput of collecting the feature data of graph nodes (usually more than 150,000) in a batch of size 1024 under \textsf{GraphSage} with 2-layer sampling. We compare \sys with a state-of-the-art RPC library, \emph{TensorPipe}~\cite{tensorpipe}, which is the high-performance NCCL-backed RPC library of PyTorch, as used by DGL. We deploy the experiment on the \textsf{cluster testbed} with 3 servers interconnected by InfiniBand. Each pair of GPUs uses NVLink.
For the \textsf{paper100M} dataset (see~\F\ref{fig:unified-tensor-paper}), the RPC library collects features at the rate of 3\unit{GB/s}, but \sys{}'s feature collection achieves 7\unit{GB/s} using NVLink. With InfiniBand, avoiding the slower Ethernet links, \sys reaches 18\unit{GB/s}. Since \sys can leverage both NVLink and InfiniBand, it achieves a combined throughput up to 40\unit{GB/s}, which is 13$\times$ higher than that of the RPC library.
We observe a similar performance improvement for the larger dataset, such as \textsf{mag240M} (see~\F\ref{fig:unified-tensor-mag}). \sys{}'s high feature collection throughput shows the benefit of using GPUs for feature aggregation together with one-sided reads that employ CPU by-pass, surpassing the performance of conventional approaches that coordinate GPU's collective communication through CPUs~\cite{kungfu2020}.

\subsection{Impact of communication links}
\label{sec:eval:comms_links}


We report \sys{}'s performance (in terms of latency) in different network configurations: \sys without InfiniBand and \sys without NVLink. Specifically, we disable InfiniBand by using SSD as the storage backend, and we disable NVLink by following the strategy shown in \F\ref{fig:partition}.
When we disable InfiniBand for the mag240m dataset, the latency grows by 1.6$\times$ from 30.2\unit{ms} to 48.9\unit{ms}. When we disable NVLink for the paper100m dataset, the latency grows by 1.5$\times$ from 27.4\unit{ms} to 41.2\unit{ms}.
 Without the faster connectivity, the communication between servers and the GPUs must involve the CPU, which is slower.

\section{Conclusions}
\label{sec:concl}


We described \sys, a new low-latency GPU-based GNN serving system that is workload-aware. \sys achieves low-latency by dynamically batching requests based on latency predictions that account for the sampled sub-graph size. Our experimental results show that \sys substantially surpasses the performance of existing distributed GNN serving systems.

\section{Acknowledgement}

We thank the constructive feedback from Wenting Shen, Ye Li, Xiaoming Qin, Lingguan Yang and Kun Zhao for improving the development of an early version of \sys.






\bibliographystyle{ACM-Reference-Format}
\bibliography{main.bib}

\end{document}